\documentclass[
  reprint, 
  amsmath,
  amssymb,
  aps,
  pra, 
  superscriptaddress, 
  longbibliography 
]{revtex4-2}

\usepackage{graphicx} 
\usepackage{dcolumn}  
\usepackage{bm}       
\usepackage{xurl}
\usepackage{hyperref} 
\usepackage{xcolor}
\usepackage[capitalise]{cleveref}
\usepackage{tikz}
\usepackage{quantikz}
\usepackage{lipsum}
\usepackage{tkz-graph}
\usetikzlibrary{hobby,backgrounds,calc,trees}
\usepackage{xcolor}
\usetikzlibrary{positioning}
\usepackage{ulem}
\usepackage{yquant}
\useyquantlanguage{groups}
\usepackage{makecell}
\usepackage{tikz,esvect,tikz-3dplot}
\usetikzlibrary{3d,calc,intersections}
\usepackage{quantikz}
\usepackage{hyperref}

\usetikzlibrary{shapes.geometric, arrows}

\tikzstyle{arrow} = [thick,->,>=stealth]

\usepackage{algorithm,algpseudocode}
\algnewcommand{\To}{\textbf{To }}
\algnewcommand\Input{\item[\textbf{Input:}]}%
\algnewcommand\Output{\item[\textbf{Output:}]}%

\begin{document}

\title{Fault-tolerant and fully addressable unitary logical gates via round-robin sparsification}

\author{Kaavya Sahay}
\thanks{Current email: ksahay@mit.edu}

 \affiliation{Google Quantum AI, California, USA}

\affiliation{Department of Applied Physics,
Yale University,  New Haven, Connecticut 06520, USA}

\affiliation{Yale Quantum Institute, Yale University, New Haven, Connecticut 06511, USA}

\author{Cody Jones}
 \affiliation{Google Quantum AI, California, USA}


\begin{abstract}
For quantum low-density parity check (qLDPC) codes, the ability to manipulate individual logical units of information, i.e.\ the use of addressable gates, enables maximal utility of their efficient encoding.
We introduce explicit low-depth unitary circuits that realize fully addressable and fault-tolerant two-qubit Clifford gates for a subset of two-dimensional translationally invariant qLDPC codes. Our transversal unitaries are created by applying a geometrically structured sparsification procedure to round-robin circuits. We verify the general applicability of our protocol by proposing and numerically simulating a range of addressable CZs between toric, cyclic hypergraph-product, color, gross, and directional codes, demonstrating fault-tolerant memory-like logical error rate scaling.
\end{abstract}


\maketitle

\section{Introduction}

 Fault-tolerant quantum computers will use quantum error correction (QEC) codes to protect logical quantum information from decoherence~\cite{shor1997faulttolerantquantumcomputation}.
In order to reduce overhead, it is desirable that each code encodes multiple logical qubits. 
 For such quantum low-density parity check (qLDPC) codes~\cite{Breuckmann2021ldpc}, the ability to manipulate individual logical units of information, or in other words, the use of \textit{addressable} gates, enables maximal utility of their efficient encoding.

Designing fault-tolerant (FT) addressable gates is difficult.
When available, the most intuitive and low-overhead strategy for computation, \textit{transversal} gates, often act on all the logical qubits of a QEC code. 
Note that in a transversal gate, any given qubit of a QEC code unitarily interacts with a fixed number of other qubits in order to achieve a logical operation.
The bottleneck for low-overhead addressable gates thus lies in the manipulation of only a specific subset of logical qubits within a QEC code, without affecting any others.

The vast majority of works proposing addressable logical computation for QEC codes are based on an alternate non-unitary gate strategy referred to as code surgery~\cite{cohen2022lowoverheadckbb,williamson2026lowoverheadgauging,swaroop2025universaladaptersquantumldpc,cowtan2024ssipautomatedsurgery}.
In code surgery, products of logical operators are measured in order to advance logical computation.
While well-developed, surgery methods have several drawbacks.
In order to be fault-tolerant, most of these logical measurements take $O(d)$ measurement rounds.
Additionally, measurements are generally the slowest and more error-prone components of several hardware platforms~\cite{bluvstein2025fault,atomcomputing2026toric, google2021exp}.
Despite these bottlenecks, one reason for this focus on surgery is because it has not fully been understood how to perform addressable transversal gates on general qLDPC codes.

Addressable transversal gates have only recently garnered interest. 
Specifically, in Ref.~\cite{pecorari2025addressablegate}, an addressable gate for hypergraph product codes was derived by use of an auxiliary Bacon-Shor code patch. Refs.~\cite{bauer2026findingdiagonal,holmes2026quantumlogiccodes,albert2026transversality,benhemou2026automatedlogical}  investigate code families supporting low depth addressable logical unitaries. Finally, Refs.~\cite{benhemou2026automatedlogical,sahay2026computingqldpccodesclimbing} derive the action of these logical gates using chain maps, explained in \cref{app:chainmap}. 
However, these works are primarily theoretical proposals~\cite{bauer2026findingdiagonal,sahay2026computingqldpccodesclimbing}, and derive concrete transversal circuits only for a narrow subclass of codes, limited in code distance~\cite{benhemou2026automatedlogical} or stabilizer structure~\cite{sahay2026computingqldpccodesclimbing,holmes2026quantumlogiccodes,albert2026transversality,benhemou2026automatedlogical}.

In this work, we introduce structured circuits that realize addressable and fault-tolerant transversal two-qubit Clifford gates for a broad subclass of two-dimensional translationally invariant (2DTI) qLDPC codes. 
Our strategy is based on applying a geometrically derived \textit{sparsification} procedure to round-robin (RR) circuits~\cite{yoder2016universal}. This sparsification, which involves the conjugation of the original RR circuit with stabilizer deformations, can render the originally non-FT  circuits fault-tolerant.

We demonstrate our technique by proposing and numerically simulating a range of addressable CZs between 2DTI codes of arbitrary distance that include toric, cyclic hypergraph-product, color, gross, and directional codes.
Further, we show that our inter-code CZ circuits can be compiled to perform intra-code addressable CZs and simultaneous CZs between multiple logical qubits. 
Importantly, beyond theoretical fault-tolerance, we numerically and analytically study the ability of these circuits to maintain full code distance. 

Our sparsification procedure relies on the identification of multiple logical representatives of the logical qubits participating in the CZ gate, onto which the original circuit is deformed.
We achieve identification of representatives of balanced weight using the \textit{cylinder trick}, introduced in Ref.~\cite{sahay2026matchingdecoder}. 
We note that the generalization to CX gates is trivial for CSS codes, where stabilizers are composed of exclusively $X$ or $Z$ Pauli operators. 
Consequently, our approach gives rise to addressable and fault-tolerant unitary CZ and CX gates for this subclass of 2DTI codes.

The remainder of this work is structured as follows: In \cref{sec:sparsetoric}, we lay out the sparsification procedure for the simplest possible example of the toric code, producing a depth-3 circuit. Subsequently, in \cref{sec:2dti}, we describe the more complex 2DTI codes we consider, including detailing the extraction of multiple logical representatives using the cylinder trick,   and laying out a modified sparsification procedure to obtain addressable unitary CZs (tCZs). We then examine the depth, performance and fault-tolerance of our circuits in \cref{sec:results}.

\section{Sparsified Round Robin Circuits} \label{sec:sparsetoric}

\begin{figure*}
    \centering
    \includegraphics[width=0.98\linewidth]{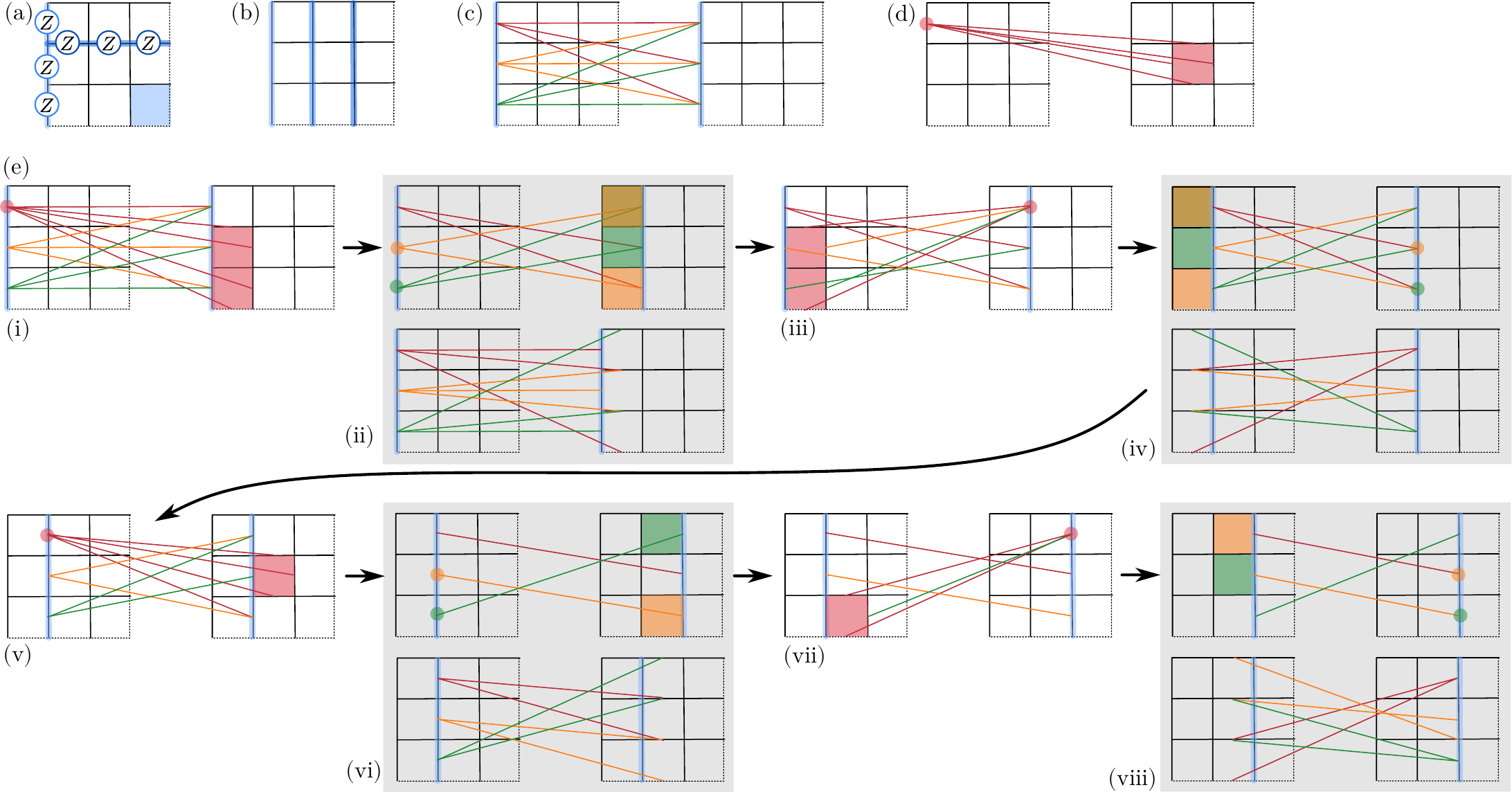}
    \caption{\textbf{A fault-tolerant addressable CZ between two toric codes} via round-robin sparsification. (a) The vertical and horizontal $Z$-logical operators of the two logical qubits of a distance $d=3$ toric code with qubits on edges (dashed lines are identified with the opposite edges). On the lower right, we show a $Z$-stabilizer.
    (b) Multiple representatives of the vertical $Z$-logical operator. 
    (c) An RR logical CZ circuit, with individual CZ gates shown by colored lines. 
    (d) A stabilizer deformation, with the stabilizer highlighted, equivalent to a logical identity. 
    (e) The sparsification procedure, consisting of conjugation of the circuit in (c) with various highlighted stabilizer deformations. Where gates become too dense within a subfigure, we show their distribution across two panels (grey shading), with the lower panel assumed to carry across subsequent steps. }
    \label{fig:toricsparsify}
\end{figure*}

We begin this section by defining round robin (RR) circuits~\cite{yoder2016universal}, which can be created between sets of physical qubits.
Consider qubit sets $\Lambda_1$ and $\Lambda_2$; a RR CZ circuit between them is:
\begin{equation}
   {RR}_{CZ}(\Lambda_1, \Lambda_2) =  \prod_{\forall i\in \Lambda_1} \prod_{\forall j\in \Lambda_2} C_iZ_{j}. \label{eq:origrrcz}
\end{equation}
Let us now restrict the qubit sets $\Lambda_1 \in \mathcal{C}_\alpha$,  $\Lambda_2 \in \mathcal{C}_\alpha$ for  two arbitrary QEC codes $\mathcal{C}_\alpha$ and  $\mathcal{C}_\beta$. We place these RR CZ circuits in the context of their action on the codespace. We provide visualizations in \cref{fig:toricsparsify} for reference, where $\mathcal{C}_\alpha, \mathcal{C}_\beta$ are both taken to be $[[2d^2, 2, d=3]]$ toric codes.

\textit{Case I:} If both $\Lambda_1$  and $\Lambda_2$  form the support of $Z$-logical operators of logical qubits $\alpha_0 
\in \mathcal{C}_\alpha$ and  $\beta_0 \in \mathcal{C}_\beta$ respectively [\cref{fig:toricsparsify}(a-b) show such logical operators for the toric code], the RR circuit is a logical CZ between $\alpha_0$  and $\beta_0$. An example logical RR circuit is shown in \cref{fig:toricsparsify}(c), where each colored line represents a gate.

\textit{Case II:} If at least one of $\Lambda_1$  or $\Lambda_2$  forms the support of a code $Z$-stabilizer on their respective codes, the RR circuit is a logical identity. We refer to RR circuits where this is true as stabilizer deformations. An example is shown in \cref{fig:toricsparsify}(d).

\textit{Case III:} This covers RR circuits where $\Lambda_1$  and $\Lambda_2$  do not satisfy Cases I or II. These circuits take the QEC system outside of its codespace.

We defer proofs that the above circuits indeed realize the stated action to \cref{sec:proof}. 
Importantly, Case I RR logical circuits are not naturally fault-tolerant. This is because any single qubit in a RR circuit for a distance-$d$ code is in the support of at least $d$ gates. We can see this using the decomposition:
\begin{equation}
   {RR}_{CZ}(\Lambda_1, \Lambda_2) =  \prod_{\forall q\in \Lambda_1} RR(q, \Lambda_2),
\end{equation}
where the RR circuit $RR(q, \Lambda_2 = \text{supp}(\overline{Z}_{\beta_0}))$ \textit{pivoted} on $q$ must have at least $d$  gates. Consequently, a single error on $q$ may spread to $\geq d $ locations via the RR circuit.

In Ref.~\cite{yoder2016universal}, these circuits were made \textit{pieceably} fault-tolerant via execution in stages separated by stabilizer measurements. 
Here, in contrast, we use stabilizer deformations to move the RR circuits off the support of a single logical representative and into the bulk of the code, i.e.\ we \textit{sparsify} the original RR circuit. We explain these stabilizer deformations using the example of the toric code, shown in \cref{fig:toricsparsify}(e).

\subsection{Toric code sparsification} \label{sec:toricsparsify}

We now go through the displayed procedure in \cref{fig:toricsparsify}(e) step-by-step. We start with an RR CZ circuit $U$ between the logical qubits $\alpha_0$ and $\beta_0$ of toric codes, $\mathcal{C}_\alpha$ and  $\mathcal{C}_\beta$ respectively. 
For the toric code, any non-trivial closed loop of $Z$s along vertical edges $v$ down the lattice  forms a representative of the same $Z$ logical operator. 
To be specific about the qubits contained in our RR logical circuits, we use the notation $\overline{Z}_{\alpha_0} = L^Z_{\alpha(v,*,i)},~ i \in \{0,1, \cdots, d-1\}$ to represent a logical $Z$ operator supported on the vertical qubits $v$ on all rows of column $i$.
We will continue to use subscripts to address lattice objects, such that $q_{\alpha(l,i,j)}$ denotes the object $q$ supported on lattice element (set) $l$, row(s) $i$ and column(s) $j$ of code $\mathcal{C}_\alpha$.
We will promote these lattice objects to corresponding Pauli operators on their support via $X,Z$ superscripts. 
With this notation, we can denote the original circuit $U$:
\begin{equation} \label{eq:sq0}
    U = RR_{CZ} (L_{\alpha(v,*,0)}, L_{\beta(v,*,0)})
\end{equation}

Our deformation strategy is highly geometrical: starting from $U$, which has support on a qubit $q_{\alpha(v,i,0)} \in L_{\alpha(v,*,0)}$, we multiply it with $RR(q, S_q)$, where $S_q^Z$ is an element of the Z stabilizer group defined by:
\begin{equation}
    S^Z_{q_{\alpha(v,i,0)}} = L^Z_{\beta(v,*,0)} \cdot L^Z_{\beta(v,*,1)} \cdot s^Z_{\beta(\square,i,0)}
    \label{eq:toricadef}
\end{equation} 
Here, $s^Z_{\beta(\square,i,0)}$ is the $Z$-stabilizer generator on a plaquette in $\mathcal{C}_\beta$ in the same row as $q_{\alpha(v,i,0)}$, between $L_{\beta(v,*,0)}$ and $L_{\beta(v,*,1)}$.
$S_{q_{\alpha(v,0,0)}}$ is shown highlighted in red in  \cref{fig:toricsparsify}(e)(i). After applying the identity CZ$^2=I$ to cancel CZ gates, \cref{fig:toricsparsify}(e)(i) also displays 
the resultant form of $U$ after deformation by $RR(q,S_q)$.
Note that $S_q$ is in the stabilizer group of $\mathcal{C}_\beta$: it contains the product of two equivalent logicals, in addition to a stabilizer generator. Thus,  $RR(q, S_q)$ forms a valid stabilizer deformation pivoted on $q$, with the stabilizer element $S_{q_{\alpha(v,i,0)}}$ spanning most of the vertical height of the lattice.

We then proceed down the first column in $\mathcal{C}_\alpha$ [\cref{fig:toricsparsify}(e)(ii)], multiplying in the corresponding $RR(q,S_q) \forall q \in L_{\alpha(v,*,0)}$ such that:
\begin{equation}
    U \gets U \cdot \prod_{q \in L_{\alpha(v,*,0)}} RR(q, S_q)
\end{equation}
As a result of these conjugations, we have pushed a majority of the support of the original RR circuit off the first column of qubits $L_{\beta(v,*,0)}$ in $\mathcal{C}_\beta$, leaving only one gate per qubit.
In other words, we have sparsified the support of the RR circuit on $L_{\beta(v,*,0)}$.
Additionally, we have accumulated two gates each on $\Lambda_{\beta(h,*,0)}$, the set of horizontal edges $h$ of $\mathcal{C}_\beta$ between the first two columns.
The reduction of gate density on $L_{\beta(v,*,0)}$ has come at the expense of increasing it on $L_{\alpha(v,*,0)}$, where each qubit now supports $d+2$ gates.

In order to move support of the circuit off $L_{\alpha(v,*,0)}$,  we can now do the same procedure on $\mathcal{C}_\alpha$. In other words,  
\begin{equation}
    U \gets U \cdot \prod_{q \in L_{\beta(v,*,1)}} RR(q, S_q),
\end{equation}
where $S_{q_{\beta(v,i,1)}} $ is defined by swapping $\alpha \leftrightarrow \beta$ in the right-hand side of \cref{eq:sq0}. This is shown in \cref{fig:toricsparsify}(e)(iii-iv), where the gates in the top panel of \cref{fig:toricsparsify}(e)(ii) are multiplied away, while those in its bottom panel have remained undisturbed. 
These steps reduce the gate depth on $L_{\alpha(v,*,0)}$ to 3, and create two more gates on each $q_{\beta(v,i,1)} \forall i$. 

At this point, we can break up the complete operation $U$ into two parts: one is the already sparsified circuit on qubit sets $\Lambda_{{\{\alpha,\beta\}}(\{v,h\},*,0)} $, displayed across the bottom panels of \cref{fig:toricsparsify}(e)(ii) and (iv), and the second is the remaining RR circuits supported on $L_{\beta(v,*,1)}, L_{\alpha(v,*,1)} $~[top panel of \cref{fig:toricsparsify}(e)(iv)]. 
Note that both codes no longer support a complete logical RR circuit $  RR_{CZ} (L_{\alpha(v,*,1)}, L_{\beta(v,*,1)}) $, but the un-sparsified part of $U$ supported on column 1 can still be broken down into smaller RR circuits supported on individual qubits.
Importantly, each component subcircuit $RR(q, \Lambda)$ now has only $(d-1)$ gates instead of $d$.

For subsequent columns $j=1,2,\cdots,d-1$, we now add in further sets of stabilizer deformations $RR(q,S_q)$ pivoted on qubits in $\mathcal{C}_\alpha$ followed by deformations pivoted on qubits in $\mathcal{C}_\beta$. These deformations sparsify support on these columns and take the form:
\begin{equation}
    S_{q_{\alpha(v,i,j)}} = L_{\beta(v,*,j)} \cdot L_{\beta(v,*,j+1)} \cdot \prod_{i' \in i-j,...,i} s^Z_{\beta^(\square,i',j)} \label{eq:toriccadef}
\end{equation} 
\begin{equation}
    S_{q_{\beta(v,i,j+1)}} = L_{\alpha(v,*,j)} \cdot L_{\alpha(v,*,j+1)} \cdot \prod_{i' \in i,...,i+j} s^Z_{\alpha(\square,i',j)}\label{eq:toriccbdef}
\end{equation} 
Note that these definitions generalize \cref{eq:toricadef} to $j> 0$. Specifically, for every subsequent column, the stabilizer element used in the deformation gets shorter -- it is shortened starting from the bottom when deforming on $\mathcal{C}_\beta$, and from the top when deforming on $\mathcal{C}_\alpha$.  
\cref{fig:toricsparsify}(d)(v-viii) shows these steps for the $d=3$ toric code.
Note that the distribution of gates shown in the bottom panels of each even step carries over across subsequent deformations. 

In general, we find that after sparsification across $(d-1)$ columns, we reach a maximal gate depth of 3 across all qubits of the toric code, independent of code size.
Consequently, this sparsified RR CZ circuit is \textit{transversal}: every qubit interacts with a constant number of other qubits to perform a logical operation~\footnote{This is a less strict interpretation of transversality as defined in Ref.~\cite{eastin2009transversal}, and aligns more with the definition of Dehn-twists in Ref.~\cite{Lavasani2019dehn}.}. 
Further, while the explanation we have provided here is sequential, since CZ gates commute, the complete tCZ circuit can be realized in any gate order that minimizes depth and implementation complexity.
For reconfigurable systems such as neutral atoms, it is desirable to understand the structure of atom rearrangements that can realize this addressable transversal CZ (tCZ). We lay out one such sequence of 3 gate layers in \cref{app:movements}.

\subsection{Fault-tolerance}

\begin{figure}
    \centering
    \includegraphics[width=0.98\linewidth]{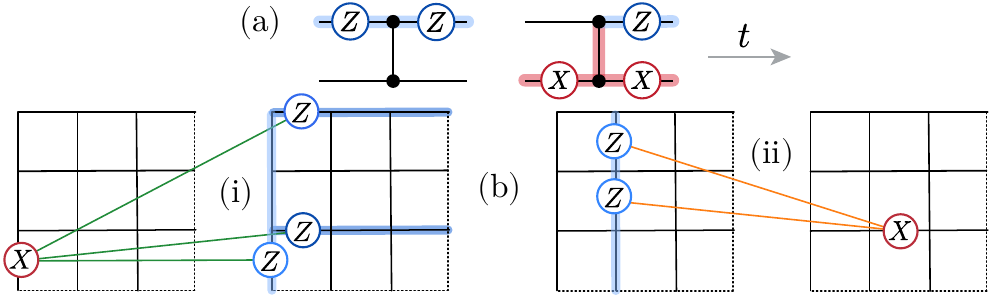}
    \caption{\textbf{Propagation of faults through a toric tCZ.} (a) Error propagation rules through a CZ gate. (b) $X$ errors propagate to $Z$s in two distinct ways: (i) Weight 3 but anti-aligned with least-weight logicals, which natively preserves the distance, or (ii) Weight-2 along a logical. The original $X$ error creates additional detection events which can identify this mechanism without reducing distance. }
    \label{fig:toricft}
\end{figure}

We can show that our tCZ construction has a \textit{fault distance} of $d$ on the toric code. In other words, the sparsified round-robin circuit requires exactly $d$ independent error mechanisms to occur in order to create an undetectable logical error.
In order to determine toric tCZ fault-tolerance, let us first establish which errors might be \textit{contagious} within our tCZ circuit, i.e.\ spread from one qubit to many: 
Since a CZ gate only spreads $X \rightarrow Z $ errors [\cref{fig:toricft}(a)], only $X$ errors may be contagious.

We find that qubit connectivity between $\mathcal{C}_\alpha$ and $\mathcal{C}_\beta$ can be classified into two subsets: For one subset, shown in \cref{fig:toricft}(b)(i), the qubit connectivity is anti-aligned with the logical operators of the code.
Here, propagation of $X$ errors is trivially fault-distance preserving, as no minimum-weight $Z$ logical can receive $>1$ error. 
    For the other connectivity subset [\cref{fig:toricft}(b)(ii)], a single $X$ error may create two $Z$'s along a logical; however, detection of the original $X$ error can help identify this mechanism: it would take the $X$ error flipping back to reduce the detection event set to that of a bare weight-2 error along the logical. As a result of these extra detection events, fault distance is maintained.  
Note that if the code to which errors were propagated was considered in isolation, this mechanism would indeed reduce the distance.
We discuss fault-tolerance for more general codes in \cref{sec:results}.

\section{2DTI codes} \label{sec:2dti}

The demonstrated sparsification procedure for the toric code allows us to lay out an algorithm for round-robin sparsification that can be applied to more general two-dimensional translationally invariant (2DTI) codes, which have been shown to reduce to copies of the toric code~\cite{chen2010localunitary,bombin2012universal,bombin2014structure,tan2026generalizedmatching,wang2026decoupling2d,chen2025anyon}. Here, we provide relevant definitions for the 2DTI codes investigated in this work. 
Our definitions are intentionally minimal and geometrical; we refer the reader to Refs.~\cite{bravyi2024highthreshold,chen2025anyon,liang2025generalized} for exhaustive algebraically-motivated descriptions of these codes. 

\begin{figure*}
    \centering
    \includegraphics[width=0.9\linewidth]{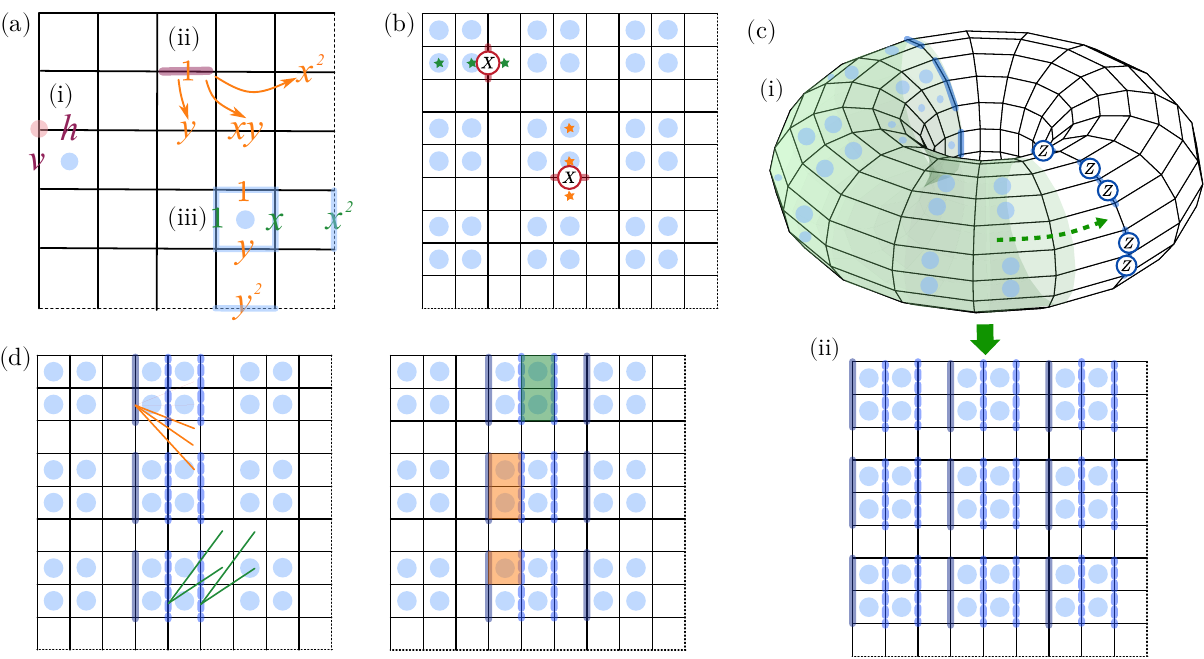}
    \caption{\textbf{Using the cylinder trick to extract disjoint logical representatives for RR sparsification on 2DTI codes.} (a) Defining lattice elements: (i) edges support $h$ and $v$ qubits (purple); $X$ (red) and $Z$ (blue) ancillae are on vertices $\bullet$ and plaquettes $\square$ respectively. (ii) Monomials denote cardinal translations across the lattice. (iii) Stabilizers can be defined via polynomials, shown for cHGP2. (b) $Z$-Symmetries of 2DTI codes are sets of checks that always product to identity: Any $X$ error leaves an even parity of syndromes in the support of the cHGP2 symmetry.
    (c) (i) Using the cylinder trick on a cHGP2 symmetry enables the extraction of a $Z$-logical operator. (ii) Sliding this selection cylinder enables extraction of $O(d)$ equivalent logical representatives. Each representative is shown via a dashed bright or solid dark blue highlight; note that certain representatives span two columns.
    (d) We show the stabilizer elements $S_q$ for specific cHGP2 qubits, derived from the rules in \cref{eqn:alg2dti}.}
    \label{fig:2dtidefs}
\end{figure*}

We consider 2DTI codes supported on a square lattice with periodic boundary conditions.
Each lattice site supports four elements: a qubit $h$ on the horizontal edge, a second qubit $v$ on the vertical edge to its immediate southwest, an $X$ ancilla on the vertex $\bullet$ between $h$ and $v$, and a $Z$ ancilla on the adjacent plaquette $\square$ contained in their cup-product~\cite{algtopbook}. This is shown in \cref{fig:2dtidefs}(a)(i).
Lattice elements can be addressed via monomials [\cref{fig:2dtidefs}(a)(ii)]. Relative to a particular lattice site denoted the origin 1, elements at translations to the right can be labeled by $x, x^2, \cdots$, and translations downwards by $y, y^2, \cdots$. On a finite lattice of size $M\times N$ with periodic boundary conditions, $x^M = 1, y^N = 1$.

\begin{table}[t]
\begin{center}
\begin{tabular}{|l|c|c|l|}
\hline
\textbf{Name} & $\boldsymbol{[[n,k,d]]}$    & 
$\boldsymbol{M,N}$ & Polynomials \\
\hline 
\hline
 Toric & $[[2l^2,2,l]]$  &   $l,l$ &  $A=1+x$ \\
 & & &  $B=1+y$ \\
 \hline
Toric color & $[[18l^2,4,3l]]$  &   $3l,3l$ &  $A=1+x+xy$ \\
 & & & $B=1+y+xy$ \\
\hline 
 Gross  & $[[288,16,12]]$ &   $12,12$ &  $A=1+x+x^{-1}y^{3}$ \\
  & & & $B=1+y+y^{-1}x^{3}$  \\
\hline
NE$^3$N  & $[[36l^2,4,4l]]$ &   $9l,2l$ & $A=1+x^3y^{-1}$ \\
 & & & $B=1+x+x^2$  \\
\hline
cHGP2& $[[18l^2,8,2l]]$ &   $3l,3l$ & $A=1+x+x^2$ \\
 & & & $B=1+y+y^2$ \\
\hline
cHGP3& $[[98l^2,18,4l]]$ &   $7l,7l$ & $A=1+x+x^3$ \\
 & & & $B=1+y+y^3$ \\
\hline
\end{tabular} 
\end{center}
\caption{Examples of relevant 2DTI codes from Refs.~\cite{geher2026directionalcodes,aydin2026cyclichypergraphproductcode,bravyi2024highthreshold}. We parameterize codes by their polynomials and lattice sizes (bound to $l \in \mathbb{N}$), which together determine $n,k,d$.}
\label{table:codes}
\end{table}

2DTI code stabilizers can be described by polynomials $A(x,y)$ and $B(x,y)$. We have
$A(x,y) = \sum_i A_i$, where each monomial $A_i = x^{a_i}y^{a_i'}$ represents a lattice site location on which the stabilizer has support. We use $\text{supp}(A)$ to denote the coordinates of the polynomial's lattice sites. The structure of 2DTI stabilizers can thus be defined as:

\begin{align}
     s^Z_{(\square, i,j)} &= q^Z_{(v,\text{supp}(x^jy^iA))} \cdot q^Z_{(h,\text{supp}(x^jy^iB))} \\
     s^X_{(\bullet, i, j)} &= q^X_{(h,\text{supp}(x^jy^iA^T))} \cdot q^X_{(v,\text{supp}(x^jy^iB^T))} .
\end{align}
Here, $A^T$ denotes inversion of the polynomial such that $x \rightarrow x^{-1},y \rightarrow y^{-1}$. In \cref{fig:2dtidefs}(a)(iii), we display $s_Z$ for a 2DTI code with $A=1+x+x^2, B=1+y+y^2$, which we refer to hereon as the cyclic hypergraph product(HGP)-2 or cHGP2 code. 
As shown, the $Z$ stabilizer has Pauli-$Z$ support on vertical qubits at lattice translations $1,x, x^2$ (from $A$) and horizontal qubits at $1,y,y^2$ (from $B$). By changing lattice dimensions and polynomials, we can define a variety of codes. We list relevant codes in \cref{table:codes} with their ${[[n,k,d]]}$ encoding parameters.

\subsection{The Cylinder Trick}

The sparsification we performed in \cref{sec:toricsparsify} relies on identification of $O(d)$ disparate logical representatives that we can sequentially deform the original RR circuit onto.
It is desirable that these representatives are maximally disjoint, and have uniform weight.
Finding $d$ such representatives is intuitive for the toric code, however, it becomes more involved for other code families.
Here, we use the cylinder trick~\cite{sahay2026matchingdecoder} to extract disjoint logical representatives with balanced weight.

We first explain this trick, which relies on the identification of linearly independent $Z$-\textit{symmetries}, or global redundancies, of a code. 
A symmetry $\Sigma^Z$ is a collection of checks such that \begin{equation}
    \prod_{c\in\Sigma^Z} c = \mathbb{I} .
\end{equation}
Identification of symmetries can be done via Gaussian elimination on the code parity check matrix, further explanation of which is provided in Ref.~\cite{sahay2026matchingdecoder}. For the codes we consider, there are always $k/2$ independent $Z$-symmetries, each of which supports two logical operators respectively spanning the lattice vertically and horizontally.

On a given symmetry, we take the product of all stabilizer generators contained within a \textit{selection cylinder} on the code's toric manifold~[\cref{fig:2dtidefs}(c)(i)]. 
Without loss of generality, we take this cylinder to span from $x=0$ to $x=M/2$, half the width of the complete lattice.
Provided that this width spans more than a stabilizer's geometric radius, this product of stabilizers can be decomposed into the product of two equivalent, disjoint logical operators $L_{(\cdot,\cdot, \sim0)} ,L_{(\cdot,\cdot, \sim \frac{M}{2})}$, where we use $\cdot$ to represent some subset of $\{v,h\}$ and the lattice rows, and ${\sim}j$ to indicate that these operators may be supported on a column set in the vicinity of column $j$. 
By sliding this selection cylinder along the torus, we can isolate different representatives of the same logical operator as the portion of the product at one boundary $L_{(\cdot,\cdot, \sim j)}, j \in \{0,...M-1\}$.

\begin{figure}
    \centering
    \includegraphics[width=0.9\linewidth]{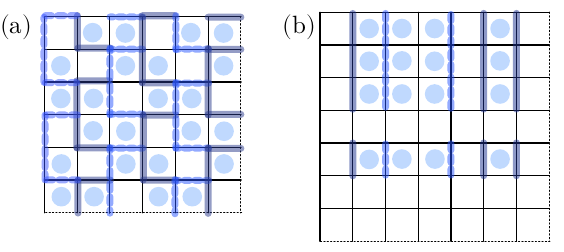}
    \caption{\textbf{$\mathcal{L}$ for more 2DTI codes.} We show the maximum disjoint logical representative set $\mathcal{L}$ of size $|J|$ obtained from the sliding cylinder trick for the (a) $d=6$ color and (b) $d=4$ cHGP3 codes. Similar to \cref{fig:2dtidefs}, adjacent disjoint logical representatives are shown via bright and dark blue highlights on the symmetry that supports them. For cHGP3, $|J| < d$. }
    \label{fig:symlogs}
\end{figure}

We show the logical representatives obtained from the cylinder trick for cHGP2 in \cref{fig:2dtidefs}(c)(ii), and the color and cHGP3 codes in \cref{fig:symlogs}.
Note that while sliding of the selection cylinder will always identify a logical representative, in order for our sparsification algorithm to work, the representatives we identify must be completely disjoint.
As a result, we do not always use all representatives, but select a maximum disjoint subset $\mathcal{L}$:
\begin{equation}
\begin{aligned}
 \mathcal{L} =
{
L_{(\cdot,\cdot,\sim j)} : j \in J
}, &
\\
L_{(\cdot,\cdot,\sim j)}
\cap
L_{(\cdot,\cdot,\sim i)}
= \emptyset
\quad \forall i,j &\in J, i \neq j,   
\end{aligned}
\end{equation}
where $J \subseteq \{0,\ldots,M-1\}$
is the set of indices corresponding to the selected representatives.
In order to find these representatives for all logical operators of a 2DTI code, we use the sliding cylinder trick oriented vertically and horizontally on a maximal linearly independent set of symmetries.

\subsection{Sparsification for 2DTI codes}

The more complex stabilizer structure of 2DTI codes in comparison to the toric code means that we have to make some modifications to the previously described sparsification procedure of \cref{sec:toricsparsify} in order to produce a short-depth and fault-tolerant circuit. Here, we describe this more general algorithm, designed for 2DTI codes with $|A|, |B| \leq 3$. Our generalization to 2DTI codes essentially swaps the strict notion of euclidean locality used for the toric code for a more relaxed notion of locality within a constant radius, determined by the structure of the code symmetry. 

Specifically, starting with $\mathcal{C}_\alpha = \mathcal{C}_\beta$ i.e. separate instances of the same code, and focusing on the corresponding logical qubits  $\alpha_0$ and $\beta_0$ supported on the same symmetry $\Sigma$,  we can define the unitary $U$:
\begin{equation} \label{eq:sq02dti}
    U = RR_{CZ} (L_{\alpha(\cdot,\cdot,\sim 0)}, L_{\beta(\cdot,\cdot,\sim 0)}).
\end{equation}
In order to sparsify $U$, we perform a sequence of deformations $U \leftarrow  U \cdot RR(q,S_q)$ pivoted on involved qubits in $\mathcal{C}_\alpha$ followed by deformations pivoted on qubits in $\mathcal{C}_\beta$. We then repeat the process for the next column included in $J$.
For each $j \in J$
 and  $ q_{\alpha(\cdot,i,j')} \in L_{\alpha(\cdot,\cdot,\sim j)}$ we have: 
\begin{equation}
\begin{aligned}
S_{q_{\alpha(\cdot,i,j')}} &=
L_{\beta(\cdot,\cdot,\sim j)} \cdot
L_{\beta(\cdot,\cdot,\sim j_{\mathrm{next}})} \cdot \\
&\prod_{\substack{j \leq k < j_{\mathrm{next}},\
i' \in {i-j,\ldots,i},\
s^Z_{\beta(\square,i',k)} \in \Sigma}}
s^Z_{\beta(\square,i',k)} .
\end{aligned} \label{eqn:alg2dti}
\end{equation}

Observe the similarity of \cref{eqn:alg2dti} to \cref{eq:toriccadef}: The stabilizer element used in any deformation contains two consecutive  logical representatives in $\mathcal{L}$, combined with a product of carefully chosen stabilizer generators between them. 
Further, this set of stabilizer generators is completely contained within the symmetry that supports $\mathcal{L}$, and expands vertically across the lattice with subsequent columns.
The difference from the toric variant is that we select stabilizer elements geometrically from the symmetry rather than the complete lattice manifold.
This is shown for specific qubits of cHGP2 in \cref{fig:2dtidefs}(d).
Similar to \cref{eq:toriccbdef}, when swapping $\alpha \leftrightarrow \beta$ indices on the right hand side of \cref{eqn:alg2dti} to create deformations pivoted on qubits $q_{\beta(\cdot,i,j'+1)}$ in $\mathcal{C}_\beta$, the condition $i' \in {i-j,\ldots,i}$ simply changes to $i' \in {i,\ldots,i+j}$, i.e. we shorten the stabilizer element in the opposite cyclic direction.

{We caution that we do not claim that this particular algorithm can directly be applied to arbitrary 2DTI codes.
Firstly, we exclusively consider codes with $|A|,|B| \leq 3$.
Further, \cref{eqn:alg2dti} relies heavily on logic borrowed from the toric code, and empirically fails to provide a low-depth circuit for certain 2DTI codes in cases when the symmetries do not possess a uniform grid-like structure. 
We note that this limitation is not indicative that full addressability is not possible via RR sparsification in these cases, but rather that \cref{eqn:alg2dti} does not use the right rules to identify $S_q$ for these symmetry structures.
Nevertheless, as we demonstrate, it has broad applicability. }

\subsection{Full addressability}
\label{ssec:fuladd2dti}
For full addressability, in addition to performing tCZ$(\alpha_0, \beta_0)$, we require the ability to perform tCZ$(\alpha_i, \beta_j) ~\forall i,j$.
Here, we explain how this is possible via the translation of code symmetries.

For {most} codes we consider, we can construct a spanning set of symmetries such that any symmetry $\Sigma_i$ is a linear translation of another, i.e.\ $\Sigma_i = x^hy^g \Sigma_j ~\forall i,j$ for some coordinates $(g,h)$.
Essentially, this means we can perform RR sparsification for vertical logical operators supported on the same $\Sigma$ for both codes, and then, dependent on the desired targets, can translate the gate supports on any code by $(g,h)$ and/or geometrically transpose them in order to address arbitrary logical operators.

\begin{figure*}[ht]
    \centering
    \includegraphics[width=0.99\linewidth]{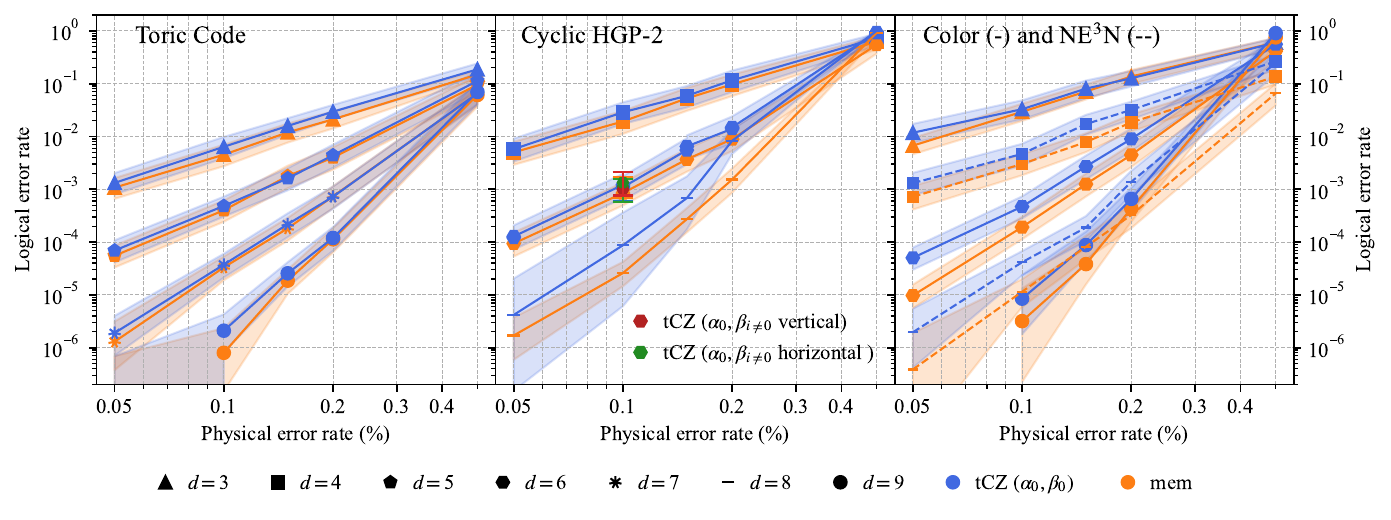}
    \caption{\textbf{Inter-code addressable tCZs  with $\mathcal{C}_\alpha =\mathcal{C}_\beta $.} We show logical error rate scaling with physical error rate for a transversal $C_{\alpha_0}Z_{\beta_0}$ (blue) compared to memory (orange) across code distances. Across panels,  $\mathcal{C}_\alpha$ is respectively defined to be toric (left),
  cyclic HGP-2 (center),
    or color and NE$^3$N directional (right) codes.
    In all plots, tCZ scaling matches that of memory.
    In the center panel, we further change the logical operator targeted in $\mathcal{C}_\beta $ via the tCZ to other vertical (red) and horizontal (red) $Z$ logical operators, showing unchanged logical performance.}
    \label{fig:3panel}
\end{figure*}

{The above strategy directly produces fully addressable tCZs for the toric, cHGP2, cHGP3, and color codes. We qualify its limits: Firstly, for a tCZ$(\alpha_i, \beta_j)$ in asymmetric codes such as the directional NE$^3$N code, when $\alpha_i$ and $\beta_j$ have orthogonal orientations, we must adjust our sparsification algorithm such that it sparsifies directly on the intended logical operators.
Secondly, when considering the gross code, we see that only a subset of symmetries can be made gridlike. 
The ability of \cref{eqn:alg2dti} to produce constant-depth addressable tCZs on the gross code is limited to the logical operators supported on these gridlike symmetries.   }

We finally note that it is not necessary that $\mathcal{C}_\alpha = \mathcal{C}_\beta$ for sparsification, and we can readily construct sparsified tCZs when $\mathcal{C}_\alpha \neq \mathcal{C}_\beta$. 
To achieve this, in principle, one may modify \cref{eqn:alg2dti} such that the elements $S_q$ are constructed respecting the symmetry structure on each code.
In practice, we use heuristics to narrow down on $S_q$ for cross-code tCZs such that we are not required to exactly extract coordinate conversion rules between dissimilar symmetry structures.

\section{Performance} \label{sec:results}

Having established an algorithmic procedure to derive tCZ circuits for 2DTI codes, in this section, we analyze the performance of our addressable tCZ constructions.

\subsection{Cross-code gate depths}

First, we derive the sparsified tCZ gate depths, i.e.\ the maximal number of gates supported on any qubit across $\mathcal{C}_\alpha, \mathcal{C}_\beta$ in our circuits. As in the toric case, our circuits have depth greater than 1.
Indeed, these tCZ circuits are not guaranteed to have minimal depth. This total depth is primarily dependent on the stabilizer structure and choice of $\mathcal{L}$, and crucially, independent of code distance.
We report these inter-code gate depths in \cref{tab:gdepth}. 

\begin{table}[h]
\caption{\label{tab:gdepth}\textbf{ Addressable tCZ gate depths for 2DTI codes.} Where a number is shown in brackets, a sparsified circuit  of the indicated lower depth was found between particular logical operators of the two codes. This is likely a result of better alignment of the directions and placement of the sparsification algorithm with these operators.}
\begin{ruledtabular}
\begin{tabular}{l|cccccc}
$\mathcal{C}_\alpha , \mathcal{C}_\beta$ & Toric & cHGP2 & Color & NE$^3$N  & cHGP3 & Gross \\
\hline
Toric & 3 & 6(5) & 6 \\
cHGP2 & & 6(5) & 8  \\
Color & & & 6 \\
NE$^3$N & & & & 8\\
cHGP3 & & & & &  14\\
Gross & & & & & & 16
\end{tabular}
\end{ruledtabular}
\end{table}
In practice, to design a sequence of CZ gates that achieves this minimal depth, we pad the gate-graph consisting of edges given by CZs with dummy edges until it is $k$-regular, and then extract maximal matchings to obtain each CZ layer. Note that using a greedy algorithm to schedule gates does not guarantee optimal depth.

\subsection{Numerical simulations}
\label{ssec:numjd}

Next, we numerically study the performance of our tCZ circuits, focusing on the resultant scaling and magnitude of logical error rate (LER) with physical error rate  $p$. 
We simulate the following system: For any given code pairing $\mathcal{C}_\alpha$ and $\mathcal{C}_\beta$ of the same distance $d$, we initialize $\mathcal{C}_\alpha$ in the $X$ basis and $\mathcal{C}_\beta$ in the $Z$ basis. 
We then perform $d$  rounds of syndrome extraction (SE) using AB sandwich SE circuits, explained in \cref{app:abswich}, such that measurements and resets for both codes are simultaneous. In the case where the SE depths for $\mathcal{C}_\alpha, \mathcal{C}_\beta$ are unequal, we insert a constant amount of idling within the SE circuit such that they are synchronized. After $d$ rounds, in the tCZ simulations, we perform the sparsified RR circuit between an arbitrary logical operator in $\mathcal{C}_\alpha$ and $\mathcal{C}_\beta$. This step is absent in the \texttt{mem} simulations used as a reference. This is followed by another $d$ rounds of syndrome extraction. Finally, we measure $\mathcal{C}_\alpha$ in the $X$ basis and $\mathcal{C}_\beta$ in the $Z$ basis.
For all circuits, we use uniform depolarizing noise, where each circuit operation on $n$ qubits experiences an error randomly chosen from $\{I,X,Y,Z\}^{\otimes n}  \backslash \{I\}^{\otimes n}$ with a probability $p$.

\begin{figure}
    \centering
    \includegraphics[width=0.9\linewidth]{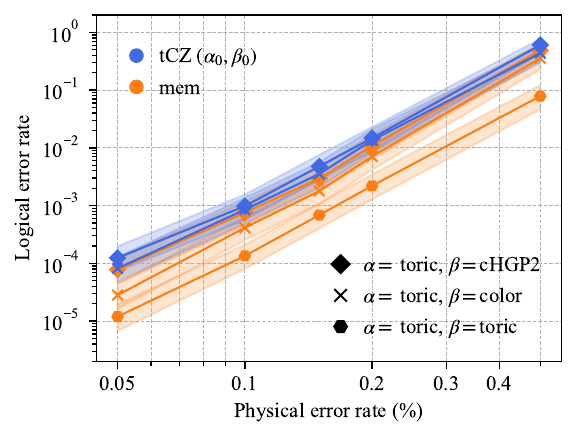}
    \caption{\textbf{Cross-code addressable tCZs  with $\mathcal{C}_\alpha \neq \mathcal{C}_\beta $.} We show LER scaling with physical error rate for a transversal $C_{\alpha_0}Z_{\beta_0}$ (blue) compared to memory (orange) for $d=6$ codes.}
    \label{fig:crosscode}
\end{figure}

We show results for various inter-code tCZ$(\alpha_0, \beta_0)$s when $\mathcal{C}_\alpha =\mathcal{C}_\beta$ in \cref{fig:3panel}, comparing our new circuits to a memory baseline \texttt{mem} for toric, cHGP2, color, and NE$^3$N codes.
{These codes are chosen such that $|J| = d$, i.e. there are $d$ disjoint logical representatives.}
In order to demonstrate full addressability, we also simulate  tCZ$(\alpha_0, \beta_i), ~\forall i\neq0$ for cHGP2. 
In all cases studied, the tCZ closely matches memory performance and LER scaling~\footnote{We observe a slight deviation between tCZ and \texttt{mem} for the $d=8$ cHGP2 at $p=0.002$. Given that we have verified that the fault distance of these circuits remains unchanged, we attribute this discrepancy to decoder suboptimality.}.
As mentioned in \cref{ssec:fuladd2dti}, we can also modify our sparsification procedure to work when $\mathcal{C}_\alpha \neq \mathcal{C}_\beta$. We show corresponding results for a subset of these codes in \cref{fig:crosscode}, once again recovering memory-like scaling for tCZs.

As a result of the tCZ, a single $X$ logical of $\mathcal{C}_\alpha$ has been conjugated with a corresponding $Z$ logical of $\mathcal{C}_\beta$. All other logical operators are invariant, up to conjugations with trivial stabilizers. We show Stim~\cite{gidney2021stim} diagrams of these logical operator transformations in \cref{app:xconjz}.

A comment on decoding: We use the `long-beam' variant of the tesseract decoder~\cite{beni2025tesseractsearchbaseddecoder}. In order to speed up decoding, we use detectors only of the relevant basis, previously done in Ref.~\cite{maan2026decoding}. Refer to \cref{app:decoders} for further analysis on how this impacts performance.
Note that for these transversal gates, it may be possible to use even more efficient decoding schemes~\cite{cain2025fastcorrelateddecodingtransversal,serraperalta2026decoding, sahay2025tcnot} that make use of the deterministic propagation of errors across the tCZ. However we leave implementation of this decoding strategy to future work. Further, where LER scaling is ambiguous, upper bounds for distances are found using the \texttt{QDistEvol} algorithm from Ref.~\cite{webster2026distancefinding} and Stim's \texttt{search\_for\_undetectable\_logical\_errors} function.

\subsection{Multi- and Intra- CZs}
\label{ssec:multiintra}

Our sparsification procedure outputs a set of gates on two codes $\mathcal{C}_\alpha$ and $\mathcal{C}_\beta$ that realize a targeted logical CZ on logical qubits $\alpha_i$ and $\beta_i$. 
In this section, we study how we can \textit{conjoin} these circuits, which we explain below.

\begin{figure}
    \centering
    \includegraphics[width=0.95\linewidth]{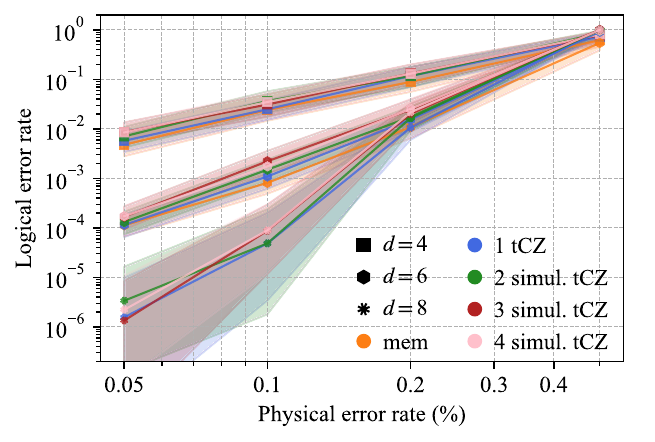}
    \caption{\textbf{Simultaneous inter-code tCZs for cHGP2}. We conduct up to 4 simultaneous tCZs by conjoining sparsified RR circuits, showing resultant memory-like LER scaling.}
    \label{fig:chgp2multi}
\end{figure}

Given gate sets for tCZ$(\alpha_0, \beta_0)$, tCZ$(\alpha_1, \beta_1)$,.. one may choose to combine these gate sets in order to simultaneously execute more than one targeted tCZ.
Interestingly, this addition  may cancel out some physical CZs in the original circuits via the identity CZ$^2=I$.
Indeed, for cHGP2, while a single tCZ has depth 6, a particular choice of two simultaneous tCZs has only depth 9, i.e. several gate layers have annihilated. Further, a conjoined circuit to conduct four simultaneous tCZs between vertical logical operators has depth 11 instead of $6\times 4 = 24$. 
Further, one may envision starting with a global transversal CZ and conjoining it with our addressable circuits in order to perform $> k/2$ simultaneous CZs in low depth.
In \cref{fig:chgp2multi} we numerically study multiple simultaneous inter-cHPG2 tCZs. As before, we recover memory-like performance.

{Another way we can conjoin circuits is within the same code to produce intra-code addressable gates. This can be done as follows: We first obtain a sparsified set of gates for tCZ$(\alpha_i, \beta_j), i \neq j$ when $\mathcal{C}_\alpha = \mathcal{C}_\beta$. Next, we take the gate support on $\mathcal{C}_\beta$ and transplant it onto $\mathcal{C}_\alpha$ in order to perform tCZ$(\alpha_i, \alpha_j)$.
Note that this increases the overall circuit depth compared to the cross-code setting.}

{On testing the above procedure for the toric and cHGP2 codes, we observe the following effects on distance: When $\alpha_i$ and $\alpha_j$ have both vertical (or horizontal) $\overline{Z}$ support, the resultant circuit has fault-distance $d/2$. When they have orthogonal orientations, the resultant circuit has fault-distance $(d-1)$. Due to the increased unitary circuit volume and the fact that we cannot filter detectors of either basis within a single patch, decoding complexity for these intra-code tCZs is notably higher than the cross-code circuits. Nevertheless, we present numerical data in \cref{app:decoders}.}

As a final remark, we do not claim that these intra- and multi-tCZ circuits  are optimal in terms of gate count or depth.  Rather, we use them as proof-of-principle demonstration of the versatility of the RR sparsification procedure in producing addressable gates.

\subsection{Fault-tolerance and more complex codes}
\label{ssec:ftchgp3}

Recall that in \cref{ssec:numjd}, we considered $|A|, |B| \leq 3$ 2DTI codes such that $d$ disjoint logical representatives were obtained from the sliding cylinder trick.  For the codes we investigated, this resulted in tCZ circuits where full memory-like fault-distance was maintained. Here, we analyze tCZs for more complex codes where $|J| < d$. We observe that using this smaller set of logical representatives leads to a constant-factor degradation in fault distance. While offering a solution to this degradation, we numerically show that this asymptotic distance reduction may not impact practical logical error rates.

When $|J| <d$, intuitively, the stabilizer element $S_q$ gets wider. Observe \cref{eqn:alg2dti}, where we include generators between $j$ and  $j_{\mathrm{next}}$ for each $ q_{\alpha(\cdot,i,j')} \in L_{\alpha(\cdot,\cdot,\sim j)}$.
The use of a geometrically wider $S_q$ increases the gate depth on each RR circuit pivoted on $q$.
Correspondingly, contagious errors on $q$ may propagate to larger regions of the lattice in the sparsified circuit, inviting distance degradation.
However, as long as the span of $S_q$ increases by no more than a constant factor, this degradation maintains asymptotic fault-tolerance.

For the cHGP3 and gross codes we consider, the maximal disjoint set $\mathcal{L}$ skips at most one consecutive logical representative, independent of lattice size. 
This approximately doubles  the geometric span of the largest $S_q$ elements. 
Correspondingly, we observe a reduction in distance from $d$ to $d/2$ in the bare cHGP3 tCZ circuit. We  remedy this distance degradation via the insertion of a single round of $Z$-stabilizer measurements within a careful ordering of the tCZ circuit, justified and described in further detail in \cref{app:ZSE}.
Consequently, we produce a \textit{distance-optimized} tCZ variant, where a carefully chosen subset of gates of the tCZ is first performed, followed by  a single $Z$-stabilizer SE round to detect contagious errors, followed by the remainder of the tCZ gates.
For the $d=4$ cHGP3 code, we verify that the distance-optimized tCZ variant restores full fault distance compared to the reduced fault distance of 2 of the naive tCZ circuit.

\begin{figure}
    \centering
    \includegraphics[width=0.85\linewidth]{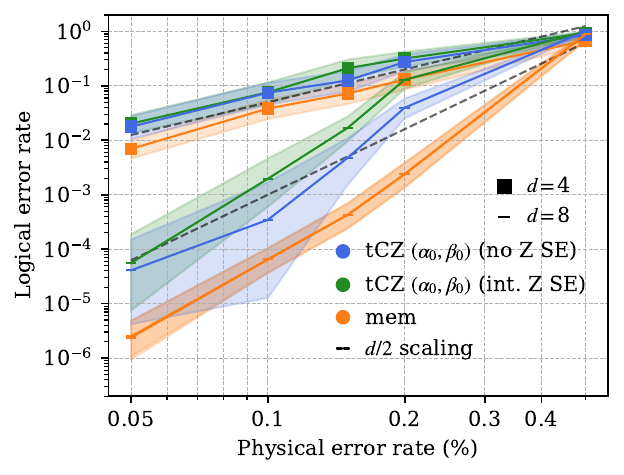}
    \caption{\textbf{Inter-cHGP3 addressable tCZs}. The logical error rates of the distance-optimized tCZ circuits with an intermediate $Z$ stabilizer measurement round restoring fault distance $d$ (green) are similar to the naive circuits with fault distance $d/2$ (blue).}
    \label{fig:chgp3}
\end{figure}

We display the results for cHGP3 tCZs in \cref{fig:chgp3}. 
Note the increase in LER for the tCZs relative to the memory baseline. 
Surprisingly, for the physical error rates we consider, the asymptotic distance degradation does not appear to impact logical error rate scaling, indicated by a dashed line fit to $O(p^{d/2})$. 
Specifically, the naive tCZ implementation displayed in blue with reduced fault distance appears to have near indistinguishable or slightly improved logical error rates compared to the distance-optimized tCZ implementation in green.
We attribute this to the combinatorial multiplicity of higher-weight errors dominating LERs for the reduced fault-distance circuits, as reported in prior works for memory experiments~\cite{gu2026scalableneuraldecoderspractical,westerheim2026walkingfloquetcodecircuits}.
Indeed, the higher overall circuit volume and increased decoding complexity of the distance-optimized tCZ variant may render it less attractive to practical implementation than the naive distance-degraded version.

\begin{figure}
    \centering
    \includegraphics[width=0.9\linewidth]{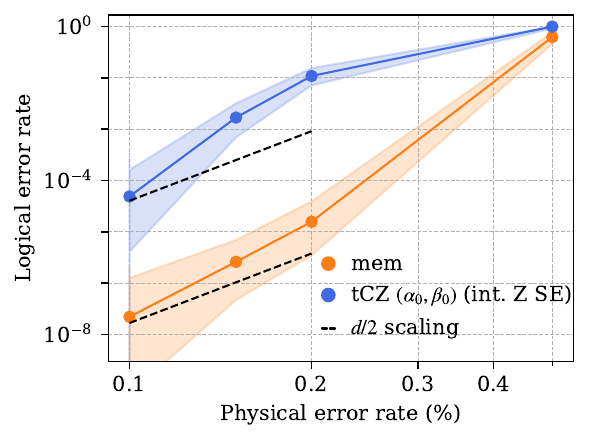}
    \caption{\textbf{An inter-gross addressable tCZ}. We simulate the 288-qubit gross code with code capacity distance $d=12$. While there is a marked difference in the LER magnitudes for the tCZ and memory experiments, the tCZ displays fault-tolerant LER scaling.}
    \label{fig:gross}
\end{figure}

{Using the distance-optimization rules of \cref{app:ZSE}, we further generate an addressable tCZ circuit for the gross code, numerical results for which are displayed in \cref{fig:gross}.
We compare this inter-gross code tCZ to code surgery, specifically an in-module $XX$ measurement from Ref.~\cite{yoder2025tourgrossmodular}, describing each circuit in more detail in \cref{app:gross}. These operations are chosen to respectively represent the simplest and best performing gross code tCZ and measurement-based primitives. On comparison, our addressable tCZ shows a ${\sim}4\times$ improvement in spacetime overhead and a ${\sim}5\times$ lower impact on logical error rate relative to code surgery.}

\section{Conclusion}

In this work, we have derived circuits to perform fully addressable unitary logical CZ gates, in circuit depth independent of code distance, for certain two-dimensional translationally invariant qLDPC codes.
We demonstrate a range of cross-code, intra-code, and simultaneous addressable CZs for the toric, color, directional, cyclic hypergraph-product and gross codes, showing memory-like logical error rate scaling.

Intuitively, we use the argument that a logical round-robin CZ circuit that initially has ${\sim}d^2$ gates can be sparsified across the $O(d^2)$ physical qubits of these 2DTI codes.
In this work, this sparsification is performed highly geometrically, using a strict notion of 2D locality.
A promising future direction is to generalize our sparsification algorithm to 2DTI codes described by polynomials of higher degree and weight.
Naturally, this raises the question of whether there are more general code families on which this sparsification can be performed.
Though we have only studied codes with periodic boundary conditions, our method is generalizable to codes with boundaries, expanded on with a basic example in \cref{sec:steane}.
In contrast, it remains unclear whether this sparsification procedure may yield FT circuits for broader code families that do not respect coarse-grained locality.

A straightforward extension of our results is the sparsification of C$^r$Z on an $r$-dimensional translationally invariant code. Specifically, we hope to find low-depth addressable CCZ circuits for analogous 3DTI codes, similar to Ref.~\cite{sahay2026computingqldpccodesclimbing}'s construction for the 3D toric code. 
Further, one may try to sparsify a C$^r$Z on an $(r-r')$-dimensional code in order to reduce circuit depth.
This procedure may enable lower-overhead magic state preparation in codes where non-Clifford gates are not transversal.

More broadly, the introduction of a plethora of addressable transversal gadgets in our work brings into question the foundations of recent architectural proposals, where transversal operations are restricted either to, for instance, the surface code~\cite{khan2026architectingearlyfaulttolerant}, or are performed in batches~\cite{xu2025batchedhighratelogicaloperations}.
Naturally, this brings into question whether these unitary circuits may be performed without $d$ rounds of intermediate error correction, as Ref.~\cite{zhou2025transversalft} suggests.
As indicated by our gross code analysis, provided that the reconfiguration cost is not insurmountable for neutral atom architectures, we expect that our unitary circuits may prove faster and easier to correct than surgery-based approaches.

\textit{AI Disclosure --} Gemini was used to iterate on maximal-matching and constraint-sliced gate scheduling, make faster bit-packed versions of human-generated GF(2) rowspace functions, and for minor  formatting.  All text and figures were produced by the authors.

\textit{Data availability -- } We provide sample circuits and data at a \href{https://github.com/kaavyas99/sparsified_tczs.git}{companion github repository}.

\textit{Acknowledgments --}
KS thanks Craig Gidney for lessons on how to use \href{https://github.com/quantumlib/Stim/tree/main/glue/stimflow}{Stimflow}. 
We are grateful to Yingjia Lin, Arshpreet Maan, Sunny He, Leonid Pryadko, Shalby M.A., Qile Su, and Oscar Higgott for useful discussions.

\bibliography{ldpc_gates}

\newpage
\appendix
 \crefalias{section}{appendix}

\section{Logical operations via Chain Maps}
\label{app:chainmap}

Here, we show that unitary logical operations, specifically CXs on CSS codes, can be recovered using the chain map formalism. Recall that a CSS code $\mathcal{C}_\alpha$ can be written as a chain complex.
\begin{equation}
    C_2^\alpha \xrightarrow{\partial_2^\alpha} C_1^\alpha \xrightarrow{\partial_1^\alpha}  C_0^\alpha ,
\end{equation}
where $C_2^\alpha$ is the vector space of $Z$-stabilizers, $C_1^\alpha$ is the space of qubits, and $C_0^\alpha$ is the space of $X$-stabilizers.
Further,  $\partial_1$ is the $H_X$ parity check matrix and $\partial_2$ is $H_Z^T$ such that 
 $ \partial_2^\alpha \cdot \partial_1^\alpha = 0 $. 
We can define a chain map between two  code instances $\mathcal{C}_\alpha$  and $\mathcal{C}_\beta$   as:
\begin{equation}
\begin{tikzcd}
  C_2^\beta \arrow[r, "\partial_2^\beta"] \arrow[d, "\gamma_2"'] 
    & C_1^\beta \arrow[r, "\partial_1^\beta"] \arrow[d, "\gamma_1"'] 
    & C_0^\beta \arrow[d, "\gamma_0"] \\
  C_2^\alpha \arrow[r, "\partial_2^\alpha"'] 
    & C_1^\alpha \arrow[r, "\partial_1^\alpha"'] 
    & C_0^\alpha,
\end{tikzcd}
\end{equation}
where $\gamma_i$ is a binary matrix of size $|C_i^{\alpha}| \times |C_i^{\beta}|$ representing a CX operation between the elements of $C_i^{\alpha},  C_i^{\beta}$. In order for the code space to be preserved via this map, we require:
\begin{equation}
\partial_2^\alpha \gamma_2 = \gamma_1 \partial_2^\beta,
\qquad
\partial_1^\alpha \gamma_1 = \gamma_0 \partial_1^\beta.
\label{eq:chain-map-genc}
\end{equation}
If we only have a set of CX gates between qubits of the two codes,  $\gamma_2=\gamma_0 = 0$. This updates \cref{eq:chain-map-genc} to:
\begin{equation}
0 = \gamma_1 \partial_2^\beta = 
\partial_1^\alpha \gamma_1 .
\label{eq:chain-map-0}
\end{equation}

At this point, the problem has reduced to finding the form of $\gamma_1$ that satisfies both equalities of \cref{eq:chain-map-0}. In order to do this, we vectorize $\gamma_1$ and solve for the nullspace of the vertically stacked matrix

\begin{equation}
    \begin{pmatrix} (\partial_2^\beta)^T \otimes I_{n_\alpha} \\ I_{n_\beta} \otimes \partial_1^\alpha \end{pmatrix} \cdot \text{vec}(\gamma_1) = 0 ,
\end{equation}

where $n_\alpha (n_\beta)$ is the number of qubits in $\mathcal{C}_\alpha (\mathcal{C}_\beta )$. Empirically, the bare form of the solutions for $\gamma_1$ consists primarily of non fault-tolerant logical RR circuits, or stabilizer deformations.

Previous work~\cite{benhemou2026automatedlogical} has attempted to find low-depth and fault-tolerant logical solutions from the nullspace  of $\gamma_1$ by formulating and solving a CP-SAT problem with these circuits as inputs.
Specifically, a sparse and shallow solution, formed by a linear combination of a non-trivial logical circuit with specific stabilizer deformations, is found by minimizing a global cost function on gate count and circuit depth.
While this approach works quite well for small codes, it quickly becomes intractable for larger codes. 
Indeed, the depth $>5$ 2DTI circuits we find through our systematic sparsification procedure for large distances are out of reach for reasonable implementations of this SAT-solving method.

\section{Proof of action of RR circuits} \label{sec:proof}

Here, we prove the action of RR CZ circuits stated in \cref{sec:sparsetoric} on a CSS code. We start by restating \cref{eq:origrrcz}:

\begin{equation}
   {RR}(\Lambda_1, \Lambda_2) =  \prod_{\forall i\in \Lambda_1} \prod_{\forall j\in \Lambda_2} C_iZ_{j}. \label{eq:rrrestate}
\end{equation}

\textbf{Lemma I:} If at least one of the qubit sets $\Lambda_1$  or $\Lambda_2$  forms the support of a code $Z$-stabilizer, the RR circuit is a logical identity. 

\textit{Proof:} Without loss of generality, assume $\Lambda_2$ is an element of the stabilizer group. We can decompose \cref{eq:rrrestate} as
\begin{equation}
   {RR}(\Lambda_1, \Lambda_2) =  \prod_{\forall q\in \Lambda_1} RR(q, \Lambda_2), \label{eq:rrdecomp}
\end{equation}
Each $RR(q, \Lambda_2)$ pivoted on $q$ is also of the desired form. We now show that the Lemma holds for these sub-circuits, and thus, by composition, on the larger RR circuit.

Non-trivial propagation of $X$ operators may happen in two directions in each sub-circuit:

\begin{enumerate}
    \item An $X$ operator can propagate from $q$ to $Z$'s on $\Lambda_2$. However, since $\Lambda_2$ is itself an element of the $Z$-stabilizer group, this propagation is trivial.
    \item $X$ operators can propagate from $\Lambda_2$ to $Z$'s on $q$. These $X$ operators originate from $X$-stabilizers or $X$-logicals, which by construction must have an even overlap with $\Lambda_2$, a $Z$ stabilizer. Hence, an even number of $X$'s always propagate to $q$, which cancel out to identity.
\end{enumerate}
Consequently, we can see that an RR circuit of this form is a logical identity.

\textbf{Lemma II:} If both $\Lambda_1$  and $\Lambda_2$  form the support of $Z$-logical operators, the RR circuit is a logical CZ.

\textit{Proof:} Since both $\Lambda_1, \Lambda_2$ are of the same form, we can analyze propagation of $X$-operators in one direction without loss of generality. We make use of the decomposition in \cref{eq:rrdecomp}. 

Consider that an $X$ operator can propagate from $q$ to $Z$'s on $\Lambda_2$. Since $\Lambda_2$ is a $Z$ logical, this applies a logical $Z$ controlled on each $q$. The $X$ operators originate from:
    \begin{enumerate}
        \item $X$-stabilizers, which have an even overlap with $q \in \Lambda_1$. In this case the $Z$ logical is applied an even number of times due to these propagations, cancelling out to identity.
        \item $X$-logicals, which must have an odd overlap with $q \in \Lambda_1$ in order to maintain logical commutation. In this case the $Z$ logical is applied a non-trivial odd number of times.
    \end{enumerate}
Consequently, we see that a non-trivial controlled-$Z$ logical is applied via the action of this RR circuit.

\begin{figure*}[t]
    \includegraphics[width=0.98\linewidth]{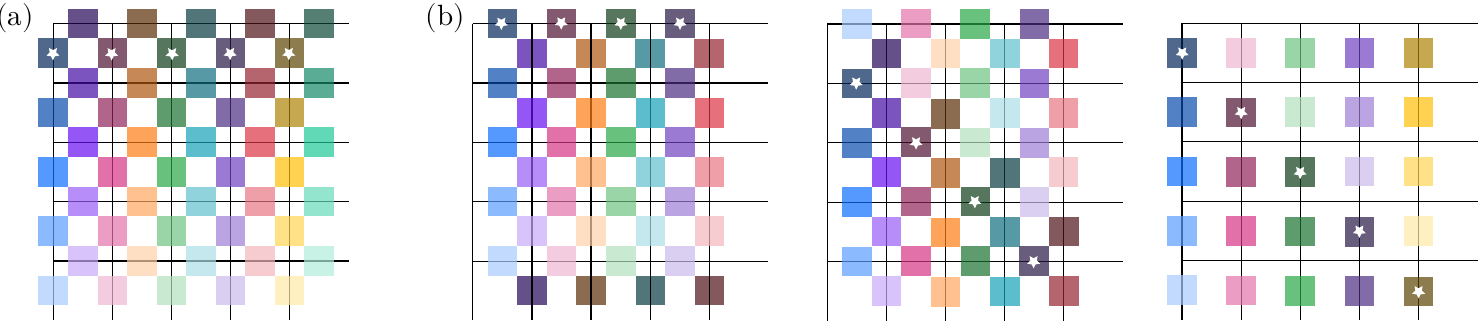}
    \caption{\textbf{Qubit positioning realizing an inter-toric addressable CZ.} (a) We show the qubits of a $d=5$ toric code via differently colored squares. For clarity, a row of qubits is highlighted using stars. (b) We show three configurations of the qubits of this toric code used for an addressable logical tCZ, realized via cyclic shifts of qubit columns; note the displacement of stars by increasing amounts across the lattice in the rightmost two configurations. When a toric code undergoes these rearrangements, with a transversal CZ performed between the displayed qubits and a second static toric code in each configuration, an addressable transversal CZ is realized. 
    } \label{fig:toricrearr}
\end{figure*}

\section{Qubit reconfigurations for inter-toric tCZ} \label{app:movements}

We show reconfiguration patterns that allow the tCZ to be performed between two $d=5$ toric codes $\mathcal{C}_\alpha$ and $\mathcal{C}_\beta$. We assume code $\mathcal{C}_\beta$ to remain static, and $\mathcal{C}_\alpha$ to be reconfigured thrice in order to realize the depth-3 logical gate. In practice, both codes may be deformed simultaneously to potentially realize the tCZ more efficiently.

We start from the original configuration of the toric code, termed $P_0$ with $v$ qubits on vertical edges and $h$ qubits on horizontal edges. 
$P_0$ is shown in \cref{fig:toricrearr}(a). Assuming a toric manifold, the first configuration $P_1$ involves no qubit rearrangement from $P_0$, but simply translation along the torus: The first four $v,h$ columns of $\mathcal{C}_\alpha$ are shifted to the northeast by $(0.5,0.5)$ lattice units and a transversal CZ is applied. Note that the last column does not experience a gate.

In the second configuration $P_2$, we perform cyclic shifts of each column of $v,h$ qubits relative to $P_0$: column $i \in \{0,1,...d-1\}$ is moved $i$ units down. The new configuration is globally moved by $(0.5, -0.5)$, and a CZ layer is performed. 
Once again, the last column does not participate in this gate layer.
In the last configuration $P_3$, we require no further reconfiguration from $P_2$, and using the global offset $(0,0)$ relative to $P_0$ we perform only the gates on the $v$ qubits.
$P_1, P_2, P_3$ are shown in \cref{fig:toricrearr}(b).

We note that the structure of this depth-3 circuit is  highly promising for practical implementation because it uses only one qubit rearrangement -- as opposed to three -- from the initial configuration. Further, this cyclic column-only reconfiguration is highly structured. 
All other movement within these three steps can be interpreted as global shifts along the torus. 
In preliminary analysis, we see that this structure can also carry forward to the tCZs we show for other codes.
Specifically, each gate layer does not need complete reconfiguration of both rows and columns from the original qubit arrangement.

\section{AB sandwich SE circuits} \label{app:abswich}

\begin{figure}
    \centering
    \includegraphics[width=0.98\linewidth]{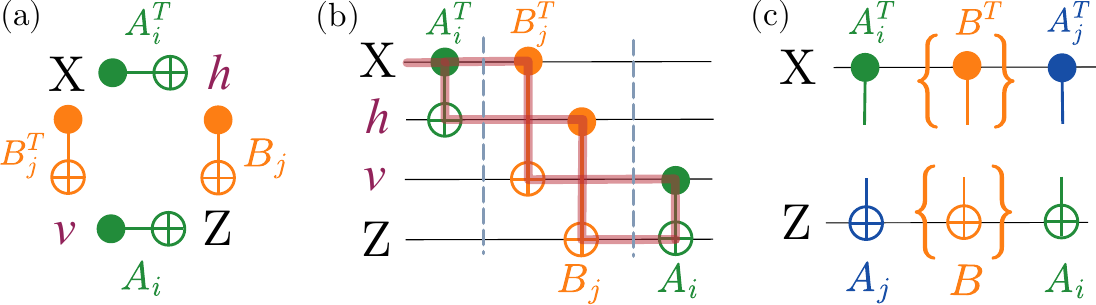}
    \caption{Generation of valid short-depth SE circuits for the codes considered in this work. (a) $X$ and $Z$ ancillas overlap on at most two qubits targeted by the same monomial bases. (b) In a valid SE circuit, cancellation of propagated X-flows from the X-ancilla on the Z-ancilla (and vice-versa) can be achieved by sandwiching the gates realizing the $B$ monomials between the $A$ monomials. (c) The monomial-pair principle from (b) can be extended to the entire polynomial, producing a valid SE circuit for  a 2DTI code.}
    \label{fig:ab}
\end{figure}

Here, we describe the syndrome extraction (SE) circuits we use for the 2DTI codes we consider in the main text. Especially for the tCZ, it is desirable that the SE rounds of $\mathcal{C}_\alpha$ and $\mathcal{C}_\beta$ are in sync such that one can efficiently stop measuring syndromes for the moments during which the tCZ is conducted.
The AB \textit{sandwich} SE circuits we describe  here achieve this goal in short depth.
This AB sandwich approach always produces a valid circuit for a weight $w$ stabilizer in depth $\leq w+1$. It is a generalization of the SE circuits proposed in Ref.~\cite{bravyi2024highthreshold}, and a special case of the SE circuits proposed in Ref.~\cite{lin2026abelianmulticyclecodessingleshot, tan26syndrome,nguyen2026depthoptimal}.
For cyclic hypergraph product codes, AB sandwich circuits are lower-depth than the cardinal variants introduced in Ref.~\cite{tremblay2022constant}.

\begin{figure*}[ht]
    \includegraphics[width=0.98\linewidth]{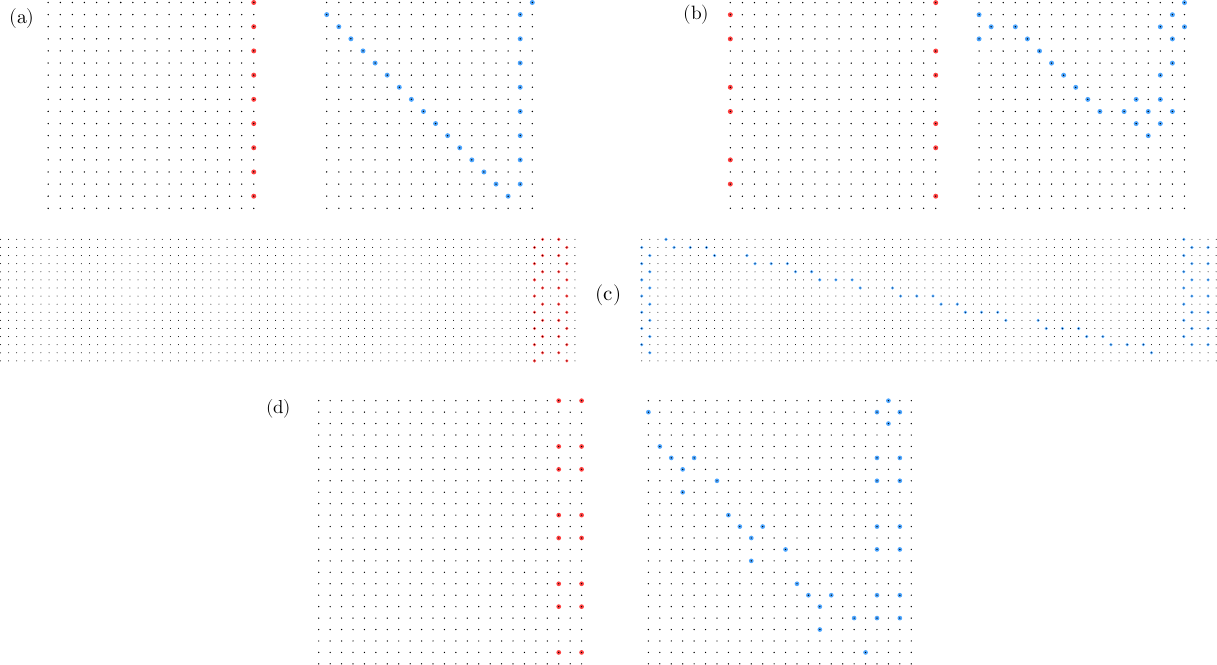}
    \caption{\textbf{Conjugated logicals through a tCZ.} We show detector-slice diagrams highlighting a specific logical operator set for the (a) $d=8$ toric code, (b) $d=12$ color code, (c) $d=16$ NE$^3$N directional code, and (d) $d=8$ cyclic HGP-2 code. For each subfigure, the displayed $X$ logical operator of the code on the left, highlighted in red, is conjugated with the shown $Z$ logical (blue) of the code on the right through the tCZ. We empirically observe the \textit{tented} logical structure laid out in Ref.~\cite{sahay2026computingqldpccodesclimbing}.}. 
     \label{fig:conjxz}
\end{figure*}

In order to produce a valid SE circuit, we must prevent entanglement of the ancillae: the flow of the $X$ operator from the X-ancilla must have trivial action on the Z-ancilla (and vice-versa). 
For 2DTI codes with $|A|, |B| \leq 3$ such as the ones we consider, $X$ and $Z$ ancillas overlap on at most two qubits targeted by the same monomial bases, say $A_i$ and $B_j$ [\cref{fig:ab}(a)]. 
In this context, a valid circuit that prevents ancilla entanglement can be produced by placing the gates realizing the $B_j$ monomial between those realizing an $A_i$ monomial, i.e. \textit{sandwiching} $B_j, B_j^T$ between $A_i, A_i^T$ [\cref{fig:ab}(b)]. 
This principle can be extended to the entire polynomials $A,B$. 
Thus, a simple way of ensuring validity is to sandwich gates realizing the full $B$ polynomial between those realizing monomials of the $A$ polynomial, or vice versa [\cref{fig:ab}(c)].

The AB sandwich circuit can always be minimal depth if at least one of $A,B$ has an even number of monomials. In this case, we sandwich the even polynomial around the odd polynomial. If both are of odd length, we are left with an extra timestep, where, say, a lone $A_i$ gate is at the beginning of the circuit, with a corresponding lone $A_i^T$ at the end.

Finally, we comment that AB sandwich circuits are not guaranteed to be optimal-depth or distance preserving: They may introduce hook errors, and for certain codes, breaking the sandwich structure might enable exactly $w$ depth circuits even when $|A|, |B|$ are odd.
What the sandwich does guarantee is short-depth valid circuits, with heuristically observed performance that is often close to  code capacity limits.
This allows us to effectively interrogate the performance of the tCZ.

\section{Form of conjugated logical operators} \label{app:xconjz}

Here, we provide Stim~\cite{gidney2021stim} detector-slice diagrams that highlight the logical $X$ operator of $\mathcal{C}_\alpha$ that is conjugated with a $Z$ logical of $\mathcal{C}_\beta$ through the tCZ. These diagrams are displayed for the toric, color, NE$^3$N, and cHGP2 codes in \cref{fig:conjxz}. We draw to the reader's attention the form of the conjugated $Z$ logical in each code. A minimum-weight $Z$ operator representative is horizontal; however, the conjugated operator is clearly not minimum-weight. Instead, it consists of qubits deformed diagonally downwards across the lattice, with a sharp turn back upwards towards the end. This form naturally follows from the structure of the sparsified circuits.

We additionally provide a live example for a tCZ on the $d=4$ toric code at the following \href{https://algassert.com/crumble#circuit=Q(0,0)0;Q(0,1)1;Q(0,2)2;Q(0,3)3;Q(0,4)4;Q(0,5)5;Q(0,6)6;Q(0,7)7;Q(1,0)8;Q(1,1)9;Q(1,2)10;Q(1,3)11;Q(1,4)12;Q(1,5)13;Q(1,6)14;Q(1,7)15;Q(2,0)16;Q(2,1)17;Q(2,2)18;Q(2,3)19;Q(2,4)20;Q(2,5)21;Q(2,6)22;Q(2,7)23;Q(3,0)24;Q(3,1)25;Q(3,2)26;Q(3,3)27;Q(3,4)28;Q(3,5)29;Q(3,6)30;Q(3,7)31;Q(4,0)32;Q(4,1)33;Q(4,2)34;Q(4,3)35;Q(4,4)36;Q(4,5)37;Q(4,6)38;Q(4,7)39;Q(5,0)40;Q(5,1)41;Q(5,2)42;Q(5,3)43;Q(5,4)44;Q(5,5)45;Q(5,6)46;Q(5,7)47;Q(6,0)48;Q(6,1)49;Q(6,2)50;Q(6,3)51;Q(6,4)52;Q(6,5)53;Q(6,6)54;Q(6,7)55;Q(7,0)56;Q(7,1)57;Q(7,2)58;Q(7,3)59;Q(7,4)60;Q(7,5)61;Q(7,6)62;Q(7,7)63;Q(16,0)64;Q(16,1)65;Q(16,2)66;Q(16,3)67;Q(16,4)68;Q(16,5)69;Q(16,6)70;Q(16,7)71;Q(17,0)72;Q(17,1)73;Q(17,2)74;Q(17,3)75;Q(17,4)76;Q(17,5)77;Q(17,6)78;Q(17,7)79;Q(18,0)80;Q(18,1)81;Q(18,2)82;Q(18,3)83;Q(18,4)84;Q(18,5)85;Q(18,6)86;Q(18,7)87;Q(19,0)88;Q(19,1)89;Q(19,2)90;Q(19,3)91;Q(19,4)92;Q(19,5)93;Q(19,6)94;Q(19,7)95;Q(20,0)96;Q(20,1)97;Q(20,2)98;Q(20,3)99;Q(20,4)100;Q(20,5)101;Q(20,6)102;Q(20,7)103;Q(21,0)104;Q(21,1)105;Q(21,2)106;Q(21,3)107;Q(21,4)108;Q(21,5)109;Q(21,6)110;Q(21,7)111;Q(22,0)112;Q(22,1)113;Q(22,2)114;Q(22,3)115;Q(22,4)116;Q(22,5)117;Q(22,6)118;Q(22,7)119;Q(23,0)120;Q(23,1)121;Q(23,2)122;Q(23,3)123;Q(23,4)124;Q(23,5)125;Q(23,6)126;Q(23,7)127;R_65_67_69_71_72_74_76_78_81_83_85_87_88_90_92_94_97_99_101_103_104_106_108_110_113_115_117_119_120_122_124_126_9_11_13_15_25_27_29_31_41_43_45_47_57_59_61_63_73_75_77_79_89_91_93_95_105_107_109_111_121_123_125_127;RX_1_3_5_7_8_10_12_14_17_19_21_23_24_26_28_30_33_35_37_39_40_42_44_46_49_51_53_55_56_58_60_62_0_2_4_6_16_18_20_22_32_34_36_38_48_50_52_54_64_66_68_70_80_82_84_86_96_98_100_102_112_114_116_118;TICK;CX_0_1_2_3_4_5_6_7_8_15_10_9_12_11_14_13_16_17_18_19_20_21_22_23_24_31_26_25_28_27_30_29_32_33_34_35_36_37_38_39_40_47_42_41_44_43_46_45_48_49_50_51_52_53_54_55_56_63_58_57_60_59_62_61_64_65_66_67_68_69_70_71_72_79_74_73_76_75_78_77_80_81_82_83_84_85_86_87_88_95_90_89_92_91_94_93_96_97_98_99_100_101_102_103_104_111_106_105_108_107_110_109_112_113_114_115_116_117_118_119_120_127_122_121_124_123_126_125;TICK;CX_0_56_1_9_2_58_3_11_4_60_5_13_6_62_7_15_16_8_17_25_18_10_19_27_20_12_21_29_22_14_23_31_32_24_33_41_34_26_35_43_36_28_37_45_38_30_39_47_48_40_49_57_50_42_51_59_52_44_53_61_54_46_55_63_64_120_65_73_66_122_67_75_68_124_69_77_70_126_71_79_80_72_81_89_82_74_83_91_84_76_85_93_86_78_87_95_96_88_97_105_98_90_99_107_100_92_101_109_102_94_103_111_112_104_113_121_114_106_115_123_116_108_117_125_118_110_119_127;TICK;CX_0_8_1_57_2_10_3_59_4_12_5_61_6_14_7_63_16_24_17_9_18_26_19_11_20_28_21_13_22_30_23_15_32_40_33_25_34_42_35_27_36_44_37_29_38_46_39_31_48_56_49_41_50_58_51_43_52_60_53_45_54_62_55_47_64_72_65_121_66_74_67_123_68_76_69_125_70_78_71_127_80_88_81_73_82_90_83_75_84_92_85_77_86_94_87_79_96_104_97_89_98_106_99_91_100_108_101_93_102_110_103_95_112_120_113_105_114_122_115_107_116_124_117_109_118_126_119_111;TICK;CX_0_7_2_1_4_3_6_5_8_9_10_11_12_13_14_15_16_23_18_17_20_19_22_21_24_25_26_27_28_29_30_31_32_39_34_33_36_35_38_37_40_41_42_43_44_45_46_47_48_55_50_49_52_51_54_53_56_57_58_59_60_61_62_63_64_71_66_65_68_67_70_69_72_73_74_75_76_77_78_79_80_87_82_81_84_83_86_85_88_89_90_91_92_93_94_95_96_103_98_97_100_99_102_101_104_105_106_107_108_109_110_111_112_119_114_113_116_115_118_117_120_121_122_123_124_125_126_127;TICK;M_9_11_13_15_25_27_29_31_41_43_45_47_57_59_61_63_73_75_77_79_89_91_93_95_105_107_109_111_121_123_125_127;MX_0_2_4_6_16_18_20_22_32_34_36_38_48_50_52_54_64_66_68_70_80_82_84_86_96_98_100_102_112_114_116_118;DT(17,1,0)rec[-48];DT(17,3,0)rec[-47];DT(17,5,0)rec[-46];DT(17,7,0)rec[-45];DT(19,1,0)rec[-44];DT(19,3,0)rec[-43];DT(19,5,0)rec[-42];DT(19,7,0)rec[-41];DT(21,1,0)rec[-40];DT(21,3,0)rec[-39];DT(21,5,0)rec[-38];DT(21,7,0)rec[-37];DT(23,1,0)rec[-36];DT(23,3,0)rec[-35];DT(23,5,0)rec[-34];DT(23,7,0)rec[-33];DT(0,0,0)rec[-32];DT(0,2,0)rec[-31];DT(0,4,0)rec[-30];DT(0,6,0)rec[-29];DT(2,0,0)rec[-28];DT(2,2,0)rec[-27];DT(2,4,0)rec[-26];DT(2,6,0)rec[-25];DT(4,0,0)rec[-24];DT(4,2,0)rec[-23];DT(4,4,0)rec[-22];DT(4,6,0)rec[-21];DT(6,0,0)rec[-20];DT(6,2,0)rec[-19];DT(6,4,0)rec[-18];DT(6,6,0)rec[-17];TICK;CZ_3_92_8_81_10_90_12_108_17_74_19_76_21_126_24_88_26_115_28_85_33_106_35_124_37_78_40_97_42_99_44_101_49_122_53_94_56_120_58_83_60_117_62_110;TICK;CZ_1_122_5_126_8_72_10_99_12_69_19_108_24_97_26_83_28_124_30_78_33_90_35_92_40_113_42_67_44_85_46_94_49_106_51_76_53_110_56_65_58_74_60_101;TICK;CZ_1_74_3_124_5_110_8_65_10_67_12_117_14_126_17_90_21_78_24_81_26_106_28_69_37_94_40_104_42_122_44_76_51_108_56_113_58_115_60_92;TICK;R_9_11_13_15_25_27_29_31_41_43_45_47_57_59_61_63_73_75_77_79_89_91_93_95_105_107_109_111_121_123_125_127;RX_0_2_4_6_16_18_20_22_32_34_36_38_48_50_52_54_64_66_68_70_80_82_84_86_96_98_100_102_112_114_116_118;TICK;CX_0_1_2_3_4_5_6_7_8_15_10_9_12_11_14_13_16_17_18_19_20_21_22_23_24_31_26_25_28_27_30_29_32_33_34_35_36_37_38_39_40_47_42_41_44_43_46_45_48_49_50_51_52_53_54_55_56_63_58_57_60_59_62_61_64_65_66_67_68_69_70_71_72_79_74_73_76_75_78_77_80_81_82_83_84_85_86_87_88_95_90_89_92_91_94_93_96_97_98_99_100_101_102_103_104_111_106_105_108_107_110_109_112_113_114_115_116_117_118_119_120_127_122_121_124_123_126_125;TICK;CX_0_56_1_9_2_58_3_11_4_60_5_13_6_62_7_15_16_8_17_25_18_10_19_27_20_12_21_29_22_14_23_31_32_24_33_41_34_26_35_43_36_28_37_45_38_30_39_47_48_40_49_57_50_42_51_59_52_44_53_61_54_46_55_63_64_120_65_73_66_122_67_75_68_124_69_77_70_126_71_79_80_72_81_89_82_74_83_91_84_76_85_93_86_78_87_95_96_88_97_105_98_90_99_107_100_92_101_109_102_94_103_111_112_104_113_121_114_106_115_123_116_108_117_125_118_110_119_127;TICK;CX_0_8_1_57_2_10_3_59_4_12_5_61_6_14_7_63_16_24_17_9_18_26_19_11_20_28_21_13_22_30_23_15_32_40_33_25_34_42_35_27_36_44_37_29_38_46_39_31_48_56_49_41_50_58_51_43_52_60_53_45_54_62_55_47_64_72_65_121_66_74_67_123_68_76_69_125_70_78_71_127_80_88_81_73_82_90_83_75_84_92_85_77_86_94_87_79_96_104_97_89_98_106_99_91_100_108_101_93_102_110_103_95_112_120_113_105_114_122_115_107_116_124_117_109_118_126_119_111;TICK;CX_0_7_2_1_4_3_6_5_8_9_10_11_12_13_14_15_16_23_18_17_20_19_22_21_24_25_26_27_28_29_30_31_32_39_34_33_36_35_38_37_40_41_42_43_44_45_46_47_48_55_50_49_52_51_54_53_56_57_58_59_60_61_62_63_64_71_66_65_68_67_70_69_72_73_74_75_76_77_78_79_80_87_82_81_84_83_86_85_88_89_90_91_92_93_94_95_96_103_98_97_100_99_102_101_104_105_106_107_108_109_110_111_112_119_114_113_116_115_118_117_120_121_122_123_124_125_126_127;TICK;M_9_11_13_15_25_27_29_31_41_43_45_47_57_59_61_63_73_75_77_79_89_91_93_95_105_107_109_111_121_123_125_127_65_67_69_71_72_74_76_78_81_83_85_87_88_90_92_94_97_99_101_103_104_106_108_110_113_115_117_119_120_122_124_126;MX_0_2_4_6_16_18_20_22_32_34_36_38_48_50_52_54_64_66_68_70_80_82_84_86_96_98_100_102_112_114_116_118_1_3_5_7_8_10_12_14_17_19_21_23_24_26_28_30_33_35_37_39_40_42_44_46_49_51_53_55_56_58_60_62;DT(1,1,1)rec[-128]_rec[-192];DT(1,3,1)rec[-127]_rec[-191];DT(1,5,1)rec[-126]_rec[-190];DT(1,7,1)rec[-125]_rec[-189];DT(3,1,1)rec[-124]_rec[-188];DT(3,3,1)rec[-123]_rec[-187];DT(3,5,1)rec[-122]_rec[-186];DT(3,7,1)rec[-121]_rec[-185];DT(5,1,1)rec[-120]_rec[-184];DT(5,3,1)rec[-119]_rec[-183];DT(5,5,1)rec[-118]_rec[-182];DT(5,7,1)rec[-117]_rec[-181];DT(7,1,1)rec[-116]_rec[-180];DT(7,3,1)rec[-115]_rec[-179];DT(7,5,1)rec[-114]_rec[-178];DT(7,7,1)rec[-113]_rec[-177];DT(17,1,1)rec[-112]_rec[-176];DT(17,3,1)rec[-111]_rec[-175];DT(17,5,1)rec[-110]_rec[-174];DT(17,7,1)rec[-109]_rec[-173];DT(19,1,1)rec[-108]_rec[-172];DT(19,3,1)rec[-107]_rec[-171];DT(19,5,1)rec[-106]_rec[-170];DT(19,7,1)rec[-105]_rec[-169];DT(21,1,1)rec[-104]_rec[-168];DT(21,3,1)rec[-103]_rec[-167];DT(21,5,1)rec[-102]_rec[-166];DT(21,7,1)rec[-101]_rec[-165];DT(23,1,1)rec[-100]_rec[-164];DT(23,3,1)rec[-99]_rec[-163];DT(23,5,1)rec[-98]_rec[-162];DT(23,7,1)rec[-97]_rec[-161];DT(16,1,1)rec[-88]_rec[-91]_rec[-92]_rec[-96]_rec[-112];DT(16,3,1)rec[-87]_rec[-90]_rec[-91]_rec[-95]_rec[-111];DT(16,5,1)rec[-86]_rec[-89]_rec[-90]_rec[-94]_rec[-110];DT(16,7,1)rec[-85]_rec[-89]_rec[-92]_rec[-93]_rec[-109];DT(18,1,1)rec[-80]_rec[-83]_rec[-84]_rec[-88]_rec[-108];DT(18,3,1)rec[-79]_rec[-82]_rec[-83]_rec[-87]_rec[-107];DT(18,5,1)rec[-78]_rec[-81]_rec[-82]_rec[-86]_rec[-106];DT(18,7,1)rec[-77]_rec[-81]_rec[-84]_rec[-85]_rec[-105];DT(20,1,1)rec[-72]_rec[-75]_rec[-76]_rec[-80]_rec[-104];DT(20,3,1)rec[-71]_rec[-74]_rec[-75]_rec[-79]_rec[-103];DT(20,5,1)rec[-70]_rec[-73]_rec[-74]_rec[-78]_rec[-102];DT(20,7,1)rec[-69]_rec[-73]_rec[-76]_rec[-77]_rec[-101];DT(22,1,1)rec[-67]_rec[-68]_rec[-72]_rec[-96]_rec[-100];DT(22,3,1)rec[-66]_rec[-67]_rec[-71]_rec[-95]_rec[-99];DT(22,5,1)rec[-65]_rec[-66]_rec[-70]_rec[-94]_rec[-98];DT(22,7,1)rec[-65]_rec[-68]_rec[-69]_rec[-93]_rec[-97];DT(0,0,1)rec[-64]_rec[-160]_rec[-164]_rec[-176];DT(0,2,1)rec[-63]_rec[-159]_rec[-163]_rec[-171];DT(0,4,1)rec[-62]_rec[-158]_rec[-162]_rec[-166];DT(0,6,1)rec[-61]_rec[-157];DT(2,0,1)rec[-60]_rec[-156]_rec[-172]_rec[-176];DT(2,2,1)rec[-59]_rec[-155]_rec[-167]_rec[-175];DT(2,4,1)rec[-58]_rec[-154]_rec[-162]_rec[-174];DT(2,6,1)rec[-57]_rec[-153];DT(4,0,1)rec[-56]_rec[-152]_rec[-168]_rec[-172];DT(4,2,1)rec[-55]_rec[-151]_rec[-163]_rec[-171];DT(4,4,1)rec[-54]_rec[-150]_rec[-170]_rec[-174];DT(4,6,1)rec[-53]_rec[-149];DT(6,0,1)rec[-52]_rec[-148]_rec[-164]_rec[-168];DT(6,2,1)rec[-51]_rec[-147]_rec[-167]_rec[-175];DT(6,4,1)rec[-50]_rec[-146]_rec[-166]_rec[-170];DT(6,6,1)rec[-49]_rec[-145];DT(16,0,1)rec[-48]_rec[-144];DT(16,2,1)rec[-47]_rec[-143]_rec[-180]_rec[-192];DT(16,4,1)rec[-46]_rec[-142]_rec[-183]_rec[-191];DT(16,6,1)rec[-45]_rec[-141]_rec[-186]_rec[-190];DT(18,0,1)rec[-44]_rec[-140];DT(18,2,1)rec[-43]_rec[-139]_rec[-188]_rec[-192];DT(18,4,1)rec[-42]_rec[-138]_rec[-179]_rec[-187];DT(18,6,1)rec[-41]_rec[-137]_rec[-182]_rec[-186];DT(20,0,1)rec[-40]_rec[-136];DT(20,2,1)rec[-39]_rec[-135]_rec[-184]_rec[-188];DT(20,4,1)rec[-38]_rec[-134]_rec[-183]_rec[-191];DT(20,6,1)rec[-37]_rec[-133]_rec[-178]_rec[-182];DT(22,0,1)rec[-36]_rec[-132];DT(22,2,1)rec[-35]_rec[-131]_rec[-180]_rec[-184];DT(22,4,1)rec[-34]_rec[-130]_rec[-179]_rec[-187];DT(22,6,1)rec[-33]_rec[-129]_rec[-178]_rec[-190];DT(1,0,1)rec[-20]_rec[-21]_rec[-24]_rec[-28]_rec[-60];DT(1,2,1)rec[-19]_rec[-23]_rec[-24]_rec[-27]_rec[-59];DT(1,4,1)rec[-18]_rec[-22]_rec[-23]_rec[-26]_rec[-58];DT(1,6,1)rec[-17]_rec[-21]_rec[-22]_rec[-25]_rec[-57];DT(3,0,1)rec[-12]_rec[-13]_rec[-16]_rec[-20]_rec[-56];DT(3,2,1)rec[-11]_rec[-15]_rec[-16]_rec[-19]_rec[-55];DT(3,4,1)rec[-10]_rec[-14]_rec[-15]_rec[-18]_rec[-54];DT(3,6,1)rec[-9]_rec[-13]_rec[-14]_rec[-17]_rec[-53];DT(0,1,1)rec[-4]_rec[-28]_rec[-29]_rec[-32]_rec[-64];DT(5,0,1)rec[-4]_rec[-5]_rec[-8]_rec[-12]_rec[-52];DT(0,3,1)rec[-3]_rec[-27]_rec[-31]_rec[-32]_rec[-63];DT(5,2,1)rec[-3]_rec[-7]_rec[-8]_rec[-11]_rec[-51];DT(0,5,1)rec[-2]_rec[-26]_rec[-30]_rec[-31]_rec[-62];DT(5,4,1)rec[-2]_rec[-6]_rec[-7]_rec[-10]_rec[-50];DT(0,7,1)rec[-1]_rec[-25]_rec[-29]_rec[-30]_rec[-61];DT(5,6,1)rec[-1]_rec[-5]_rec[-6]_rec[-9]_rec[-49];OI(0)rec[-8]_rec[-16]_rec[-24]_rec[-32];OI(1)rec[-1]_rec[-2]_rec[-3]_rec[-4]_rec[-68]_rec[-70]_rec[-71]_rec[-72]_rec[-73]_rec[-78]_rec[-82]_rec[-87]_rec[-91]_rec[-96];OI(2)rec[-93]_rec[-94]_rec[-95]_rec[-96];OI(3)rec[-68]_rec[-76]_rec[-84]_rec[-92]}{Crumble URL}. Within the Crumble interface, one may examine the propagation of all operators present within the circuit.

\begin{figure*}[t]
    \includegraphics[width=0.98\linewidth]{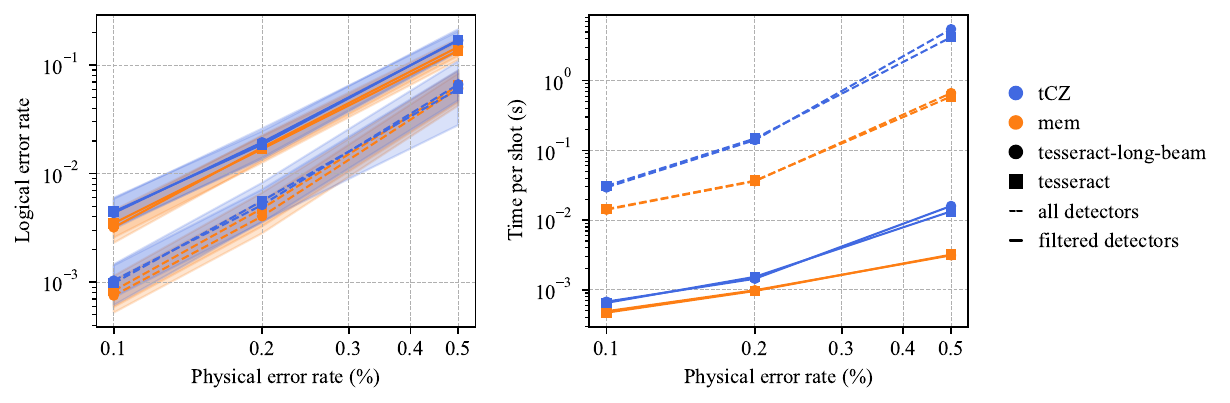}
\caption{\textbf{Benchmarking decoder variants.} We test tesseract (beam size 5) and tesseract-long-beam on the $d=4$ toric code, with and without detectors of both bases. Compared to varying beam depth, filtering detectors has a much larger impact on performance, leading to a constant factor degradation in LERs, but speedups by over an order of magnitude per shot.}
     \label{fig:decoders}
\end{figure*}

\section{Details of numerical simulations} \label{app:decoders}

In this appendix, we discuss details of the simulation parameters used to obtain results discussed in \cref{sec:results}.

\textit{Benchmarking decoders --} In \cref{fig:decoders}, we show the variation in both logical error rate and decoding time per shot across different decoder and detector set choices. 
Ultimately, in the main text, we choose a decoder variant with balanced performance on these metrics.
In particular, we may choose to filter detectors, i.e.\ only keep detectors relevant to the initialized basis of the logical qubits.
For the inter-code tCZ simulations, this involves keeping the $Z$ detectors of one logical qubit and the $X$ detectors of the other.

We see that filtering detectors leads to speedups by order(s) of magnitude, rendering investigation of our tCZ circuits practical.
Thus, for all results excluding the intra-code simulations, we use tesseract-long-beam with filtered detectors. 
For the intra-code results, we use the default shorter beam configuration of tesseract with detectors of both bases.

\textit{Distance bounds --} For most tCZ circuits, establishing tight bounds on distance proved infeasible with the methods we tested. 
Using Stim's \texttt{search\_for\_undetectable\_logical\_errors} function, we were able to obtain values that often matched asymptotic expectations with limited search degree and beam depth. 
For $d< 7$ codes, we believe these bounds may be exact given our search parameters. 
However, especially for larger distances, there is no guarantee that these bounds match the lowest weight logical error.
In this context, the scaling observed via  numerical simulations is particularly promising, as unless highlighted, we do not observe distance reduction.

\textit{Use of $d$ buffer rounds --}
For most simulations, we use $d$ buffer rounds before and after the tCZ. However, for the gross code, cHGP3, and intra-code simulations, we reduce the size of this buffer to $d/2$. 
This reduction is done in order to reduce decoding time per shot.

\textit{Intra-code tCZ --} We can readily generate intra-code tCZ circuits by conjoining the gate sets obtained from inter-code sparsification, as described in \cref{ssec:multiintra}.
However, numerically simulating these intra-patch tCZs is more involved than the inter-code case.
We make the following modifications to our inter-code simulations for an intra-code tCZ: First, we unphysically track both bases of the logical qubits in order to obtain $X$ and $Z$ failure rates. Second, we note that as the tCZ is occurring in-patch, detectors of both bases are relevant for decoding. Thus, we use both detector bases for error correction. This significantly slows down decoding.

\begin{figure}
    \centering
    \includegraphics[width=0.9\linewidth]{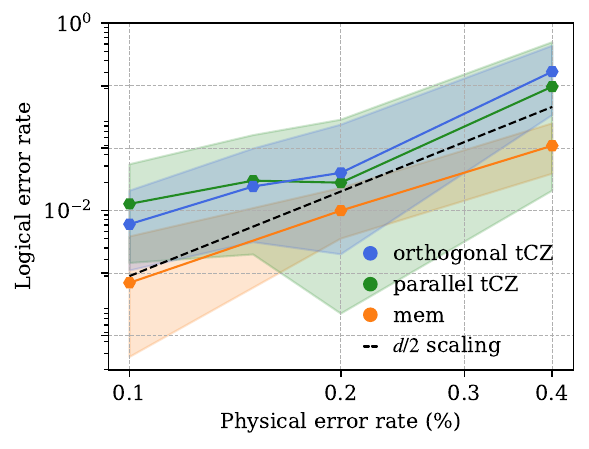}
    \caption{\textbf{Intra-code tCZs for $d=6$ cHGP2}. tCZ LERs are elevated compared to memory. However, distance-degradation is not immediately apparent in the regime shown.}
    \label{fig:intratcz}
\end{figure}

Numerical results for a tCZ$(\alpha_0, \beta_1)$ in   the $d=6$ cHGP2 code are displayed in \cref{fig:intratcz}, where the parallel (orthogonal) tCZ is conducted between aligned (perpendicular) logical operators.

\section{Using a $Z$ SE round to restore fault distance}
\label{app:ZSE}

As discussed in \cref{ssec:ftchgp3},  sparsifying on the set $\mathcal{L}$ for 2DTI codes when $|J| < d$ can reduce fault-distance by a constant factor.
We discuss remedying this distance degradation through the lens of the cHGP3 code. 
In order to do so, first recall that (1) only $X$ errors are contagious in tCZ circuits, and (2) CZ gates commute with $Z$ stabilizers, leaving them invariant throughout the depth of the tCZ circuit.

{We discuss potentially harmful error mechanisms for cHGP3,  analogous to \cref{fig:toricft}.
First, a contagious $X$ error before the tCZ may spread from a single qubit on a $h$ edge on $\mathcal{C}_\alpha$ to $Z$ errors on four $v$ qubits along a logical operator in $\mathcal{C}_\beta$. In the case another $X$ error on the first $h$ qubit flips it back to its original state, it leaves no syndromes on $\mathcal{C}_\alpha$. Consequently, this weight-2 error on $\mathcal{C}_\alpha$ has given rise to a set of syndromes consistent with a weight-4 error on $\mathcal{C}_\beta$ that is inequivalent up to stabilizers. }

We can attempt to remedy the above situation by inserting a round of $Z$ stabilizer measurements in the middle of the 4 CZ gates supported on the $h$ qubit.
We must restrict ourselves to the measurement only of $Z$ stabilizers as the $X$ stabilizers are significantly deformed mid-way through the tCZ.
Importantly, the entire tCZ circuit must be arranged, or \textit{sliced} by this $Z$ SE round, such that every qubit is connected to at most two $v$ qubits in the same column before and after this SE round. The same also applies to any qubit with CZ gates connected to $h$ qubits in the same row.
We achieve this constrained slicing of the tCZ circuit by defining an integer programming problem and inputting it into a CP-SAT solver.
In certain cases, the sliced solution has increased overall unitary circuit depth.

While performing the above slicing addresses the particular harmful error configuration described, we find that it is not always enough to restore full fault-distance for cHGP3.
We find the addition of the following constraint to be successful at accomplishing full fault-distance restoration:
For any gate set $\{(t,u_i)\}$ supported on qubit $t$, we slice the gate set such that the $X$-check support of the $\{u_i\}$ set after (before) the gate is maximally disjoint.
In other words, an $X$ error on $t$ right after the intermediate $Z$ SE round should leave the maximal possible syndrome data on $\{u_i\}$.

We finally note that the use of intermediate $Z$ stabilizers to improve the fault-tolerance of tCZs cannot directly be extended to tCXs, unlike most other facets of our tCZ construction. This is because tCXs modify the form of both $X$ and $Z$ stabilizers, and both types of errors may be contagious. Consequently, any such intermediate stabilizer measurement would have to be (1) for both bases of stabilizers, and perhaps more detrimentally, (2) performed on the potentially high weight deformed stabilizers partway through the tCX circuit.

\section{A case study on the gross code}
\label{app:gross}

In this appendix, we first discuss the design of an addressable tCZ circuit for the gross code, followed by a comparison of this tCZ with code surgery.

\begin{table*}[ht]
\centering
\begin{tabular}{|l|c|c|c|}
\hline
\textbf{} & \textbf{In-module $XX$ measurement} & \textbf{Inter-code tCZ} & \textbf{Ratio} \\ 
\hline
{SE rounds}
    & 10
    & 1  &\\
\hline
{Additional qubits}
    & 90 (bridge) + 144 ($X$ and $Z$ ancillae)
    & $72 \times 2$ ($Z$ ancillae for each code) & \\
\hline
{Total circuit depth}
    & $10 \times 9 = 90$
    & $18$ (tCZ gate layers) + $1 \times 8 = 26$
    & \\
\hline
{Total spacetime volume}
    & $90 \times (144\text{ data} + 144\text{ ancilla} + 90\text{ LPU})$
    & $18 \times (2 \times 144\text{ data})
      + (144 \times 2 + 144$ ancillae$) \times 8 $ & \\
      &  = 34020 & = 8640 & 3.94 \\
\hline
{LER degradation over }
    & & 
    &  \\
{memory at $p=10^{-3}$} & $\sim 10^{3.8}$
    & $\sim 10^{3.1}$  & 5\\
\hline
\end{tabular}
\caption{\textbf{Comparing an in-module $XX$ measurement from Ref.~\cite{yoder2025tourgrossmodular} to an inter-code addressable tCZ for the 144-qubit gross code.}
We specifically study the $X_1X_7$ measurement from Ref.~\cite{yoder2025tourgrossmodular}, which does not require further shift automorphisms, and compare it to a distance-optimized inter-code tCZ$(\alpha_0, \beta_0)$ circuit with one intermediate $Z$ SE round.
We assume a ($Z$-only) SE circuit depth of 9 (8), including measurements and resets for the deformed (original) code.   We choose not to directly compare actual LERs as Ref.~\cite{yoder2025tourgrossmodular} uses a different decoder, specifically Relay-BP~\cite{müller2025improvedbeliefpropagationsufficient}. Instead, we calculate the factor by which the LER for the logical operation being considered increases compared to the corresponding LER for memory at $p=0.001$. 
While we use a total $XX$ measurement depth of 90, Table II of Ref.~\cite{yoder2025tourgrossmodular} reports that an in-module measurement takes 120 timesteps. Using a circuit depth of 120 instead of 90 in the above calculations will increase the total measurement spacetime volume to 45360.
In practice, buffer  SE rounds may be added before and after each operation taking place within a larger quantum algorithm, affecting these numbers. }
\label{tab:xx-vs-tcz}
\end{table*}

For the gross code, the set $\mathcal{L}$ on its gridlike symmetries uses every other logical representative obtained from the sliding cylinder trick.
As a result, one might expect the same distance degradation observed in cHGP3.
To mitigate this, we input the depth-16 gate set output from RR sparsification on  $\mathcal{L}$ into the constrained slicing algorithm described in \cref{app:ZSE}.
This produces a circuit where two sets of 9 CZ layers are performed with an intermediate $Z$ SE round. Note that the effective maximum connectivity per qubit in this circuit is $16 + 6 \text{ (to $X$ and $Z$ checks)} = 22$.

Unfortunately, we are unable to find tight bounds on distance for this high-depth tCZ circuit: the upper bounds found within our search space match that of memory, but do not rule out the presence of lower-weight alternatives.
In particular, the constraints designed for distance-optimization in the cHGP3 code may not exactly solve possible distance degradation in the gross code.
Thus, we turn to numerics to verify tCZ performance.

In \cref{fig:gross}, we compare the sliced tCZ circuit to memory for the 288-qubit gross code with $d=12$.
At higher $p$, we observe a shallower scaling of tCZ LER with $p$ than desired; this might have been indicative of reduced fault distance, but given similar behaviour seen for the $d=8$ cHGP2, we attribute this to decoder ineffectiveness in a high $p$ and $d$ regime.
This effect has previously been reported in Ref.~\cite{geher2026directionalcodes}.
Indeed, at lower $p$, the tCZ LER scaling improves.

However, \cref{fig:gross} still displays significant degradation in tCZ LER compared to memory. 
It remains to be established whether this is solely a result of the higher circuit volume (there are 18 additional unitary gate layers and one depth-8 $Z$ SE round in the tCZ compared to memory), and resultant increased decoding complexity, or whether there are undiscovered lower-weight error mechanisms.
Nevertheless, given that we see fault-tolerant scaling $p^{> d/2}$, we proceed to a  comparison of this tCZ with code surgery.

In code surgery, logical operations are performed via projective multi-qubit measurements using a carefully designed ancilla system. Ref.~\cite{yoder2025tourgrossmodular} lays out a specific surgery-based architecture for the gross code family, which we use as our point of comparison.
Specifically, we benchmark against the logical primitive of an intra-code $XX$ surgery operation for the 144-qubit $d=12$ gross code as described in Ref.~\cite{yoder2025tourgrossmodular}.
While \cref{fig:gross}  uses the 288-qubit $d=12$ gross code, given that we do not expect tCZ depth to change with code size within the same code family, we can still perform a preliminary analysis.

\begin{figure*}[t]
    \includegraphics[width=0.98\linewidth]{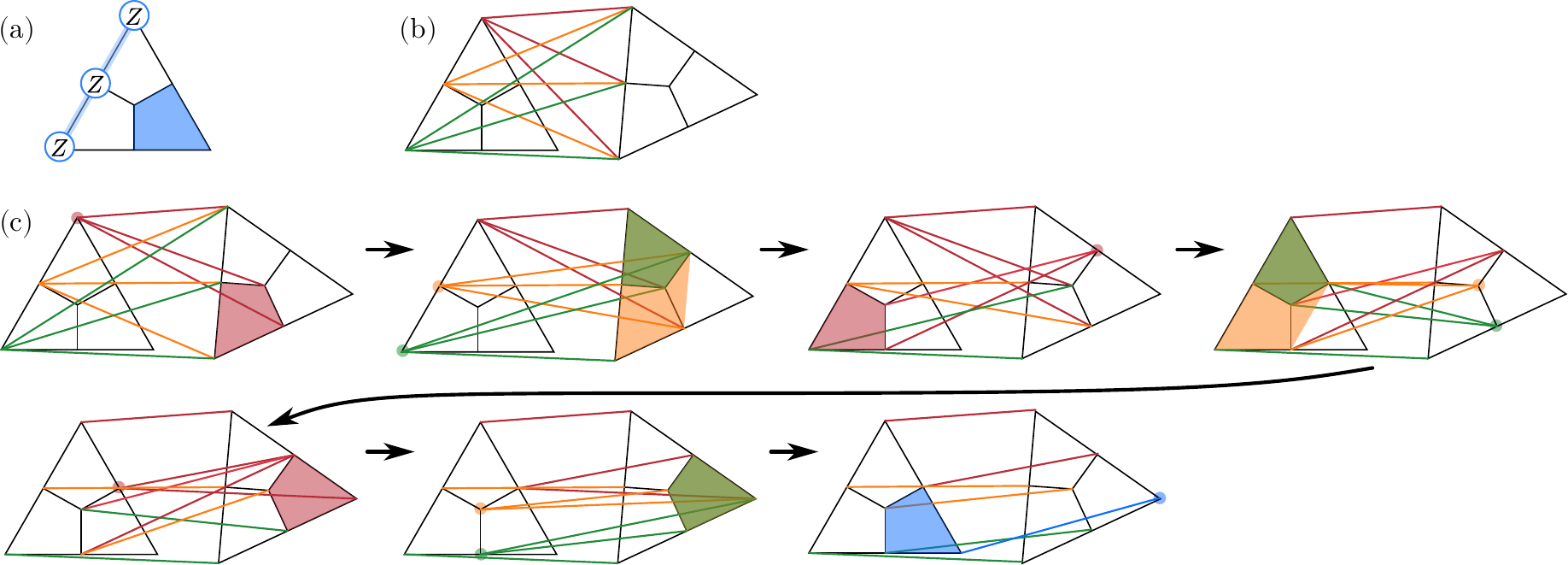}
    \caption{\textbf{RR sparsification on the Steane code yielding a depth-1 transversal CZ.} (a) A $Z$-logical operator and stabilizer, supported on the incident vertices. (b) The original RR circuit. (c) The sparsification procedure, following rules similar to \cref{fig:toricsparsify}. 
    } \label{fig:steanepush}
\end{figure*}

We note that comparing an inter-code tCZ gate to an in-code surgery operation is somewhat contrived:
For an inter-code tCZ, we benefit from a lower depth circuit and the absence of distance-reducing mechanisms reported in \cref{ssec:multiintra}. 
Further, we can reasonably obtain numerical data.
However, by virtue of operating on two code blocks, we have higher qubit overhead. Conversely, for the in-module measurement, we use lower qubit overhead, and have improved logical error rates compared to the inter-module surgery operation.
However, as both methods' best representatives, we proceed with the comparison, reporting overheads and relative LERs in \cref{tab:xx-vs-tcz}.
We observe that the tCZ improves both on the relative LERs and spacetime overhead compared to the $XX$ measurement despite acting on two code blocks.

\section{Recovering known tCZs} \label{sec:steane}

RR sparsification can lead to the recovery of well-known depth-1 transversal circuits for logical gates. 
As an example, in \cref{fig:steanepush}, we show round-robin sparsification on a RR CZ circuit between two Steane codes.
Here, sparsification yields the standard strongly transversal CZ circuit. 
This explains the decomposition in Section IV of Ref.~\cite{he2026roundrobin}.
We use the same notation as that in \cref{fig:toricsparsify}; note, however, that qubits are supported on the vertices of the lattice and $Z$-stabilizers on faces in this example. 

It is straightforward to extend the Steane sparsification procedure to any distance $d$ color code with boundaries.
Further, though we do not show it here, using the same rules on an RR CZ circuit between two rotated surface codes leads to the standard tCZ circuit in the same fashion.
We empirically observe that the presence of lower-weight boundary stabilizers simplifies the extraction of a low-depth tCZ circuit.

It is desirable to extend the sparsification procedure to more complex 2DTI codes with boundaries~\cite{nixon2026vinecodes,gu2026nearestneighbourgatesneed, steffan2025tilecodes,wang2026algebrabivariatebicycle}. Presumably, this would involve the design of a modified procedure that can extract global redundancies on a truncated lattice. The use of the cylinder trick (truncated to a \textit{rectangle} trick) on such a global $Z$-symmetry will then enable the isolation of multiple logical representatives. The RR circuit can then be sparsified on these representatives.

\end{document}